\documentclass[aps,prc,superscriptaddress,floats,floatfix,reprint,reprintnumbers]{revtex4}
\usepackage{blindtext}
\pdfoutput=1
\usepackage{epsfig}
\usepackage{graphicx}
\usepackage{dcolumn}
\usepackage{bm}
\usepackage{booktabs}
\usepackage{color}
\usepackage{mathrsfs}
\usepackage{ulem}
\usepackage{epstopdf}
\graphicspath{{plots/}}
\usepackage{verbatim}
\usepackage{indentfirst}
\usepackage{float}
\usepackage[colorlinks,
           bookmarks=true,
           linkcolor=blue,
           urlcolor=blue,
            anchorcolor=black,
            citecolor=blue]{hyperref}
\usepackage[bf,raggedright,indentafter]{titlesec}
\usepackage{rotating}
\usepackage{booktabs}
\usepackage{tabularx}
\usepackage{makecell}
\usepackage{hypcap}
\usepackage{subfigure}%
\usepackage{tabularx}

\usepackage{array}

\begin{document}
\title{\Large Semileptonic $D$ Meson Decays $D\to P/V/S\ell^+\nu_\ell$ with the SU(3) Flavor Symmetry/Breaking}
 \author{Ru-Min Wang$^{1,\dagger}$,~~~ Yue-Xin Liu$^{1}$,~~~Chong Hua$^{1}$,~~~Jin-Huan Sheng$^{2,\S}$,~~~Yuan-Guo Xu$^{1,\sharp}$\\
 $^1${\scriptsize College of Physics and Communication Electronics, Jiangxi Normal University, Nanchang, Jiangxi 330022, China}\\
 $^2${\scriptsize School of Physics and Engineering, Henan University of Science and Technology, Luoyang,   Henan 471000, China}\\
 $^\dagger${\scriptsize ruminwang@sina.com}~~
  $^\S${\scriptsize jinhuanwuli@126.com}~~
  $^\sharp${\scriptsize yuanguoxu@jxnu.edu.cn}~~
}

\begin{abstract}
Many exclusive $c\to d/s\ell^+\nu_\ell~(\ell=e,\mu,\tau)$ transitions have been well measured, and they can be used to test the theoretical calculations.
Motivated by this,  we study the $D\to P/V/S\ell^+\nu_\ell$ decays induced by  the $c\to d/s\ell^+\nu_\ell$ transitions with the SU(3) flavor symmetry approach, where $P$
denotes the pseudoscalar meson, $V$ denotes the vector meson, and $S$ denotes the scalar meson with a mass below $1$ $GeV$.
The  different decay  amplitudes of the $D\to P\ell^+\nu_\ell$, $D\to V\ell^+\nu_\ell$ or $D\to S\ell^+\nu_\ell$ decays can be related by using the SU(3) flavor symmetry and by considering the SU(3) flavor breaking.
Using the present data of $D\to P/V/S\ell^+\nu_\ell$,  we predict the not yet measured or not yet well measured processes  in  the $D\to P/V/S\ell^+\nu_\ell$ decays.
We find that the SU(3) flavor symmetry approach works well in the semileptonic $D \to P/V\ell^+\nu_\ell$ decays.  For the $D \to S\ell^+\nu_\ell$ decays,
 only the decay $D^+_s\to f_0(980)e^+\nu_e$  has been measured,  the branching ratios of the $D^+_s\to f_0(980)e^+\nu_e$ and $D\to S(S\to P_1P_2)\ell^+\nu_\ell$ decays are used to constrain the  nonperturbative  parameters and then predict not yet measured $D \to S\ell^+\nu_\ell$ decays, in addition,  the two quark and the four quark scenarios for the light scalar mesons  are analyzed. The SU(3) flavor symmetry predictions of the $D \to S\ell^+\nu_\ell$ decays need to be  further tested, and our predictions of the $D \to S\ell^+\nu_\ell$ decays are useful for  probing the structure of light scalar mesons.
Our results in this work could be used to test the SU(3) flavor symmetry approach in the semileptonic $D$ decays by the future experiments at BESIII, LHCb and BelleII.

\end{abstract}
\maketitle

\section{Introduction}
Semileptonic heavy meson decays  dominated by tree-level exchange of $W$-bosons in the standard model have attracted a lot of attention in testing the stand model and in
searching for the new physics beyond the stand model.
Many semileptonic $D\to P/V\ell^+\nu_\ell$ decays and one $D\to S\ell^+\nu_\ell$ decay  have been observed  \cite{PDG2022}, and present experimental measurements  give us an opportunity to  additionally test  theoretical approaches.

In theory, the description of semileptonic decays are relatively simple, and  the weak and strong dynamics
can be separated in these processes since leptons do not participate in the strong interaction. All the strong dynamics in
the initial and final hadrons is included in the hadronic form factors, which are important for testing the theoretical calculations of the involved strong interaction.
The form factors  of the $D$ decays  have been  calculated, for examples, by   quark model \cite{Melikhov:2000yu,Cheng:2017pcq,Soni:2018adu,Faustov:2019mqr,Chang:2020wvs,Chang:2018zjq}, QCD sum rules \cite{Ball:1993tp}, light-cone sum rules \cite{Bhattacharyya:2017wxk,Fu:2018yin,Fu:2020vqd},  covariant light-front quark models \cite{Grach:1996nz,Cheng:2003sm,Chang:2019mmh}, and  lattice QCD \cite{Lubicz:2018rfs,Lubicz:2017syv}.

The SU(3) flavor symmetry approach is independent of the detailed dynamics offering
us an opportunity to relate different decay modes,
nevertheless, it cannot determine the sizes of the amplitudes or the form factors by
itself. However, if experimental data are enough, one may use the data to extract the amplitudes or the form factors, which can be viewed
as predictions based on symmetry,  has a smaller dependency on estimated form
factors, and can provide some very useful information about the decays.   The SU(3) flavor symmetry  works well in  the $b$-hadron decays \cite{He:1998rq,He:2000ys,Fu:2003fy,Hsiao:2015iiu,He:2015fwa,He:2015fsa,Deshpande:1994ii,Gronau:1994rj,Gronau:1995hm,Shivashankara:2015cta,Zhou:2016jkv,Cheng:2014rfa,Wang:2021uzi,Wang:2020wxn},   and the $c$-hadron decays  \cite{Wang:2021uzi,Wang:2020wxn,Grossman:2012ry,Pirtskhalava:2011va,Cheng:2012xb,Savage:1989qr,Savage:1991wu,Altarelli:1975ye,Lu:2016ogy,Geng:2017esc,Geng:2018plk,Geng:2017mxn,Geng:2019bfz,Wang:2017azm,Wang:2019dls,Wang:2017gxe,Muller:2015lua}.

Semileptonic decays of $D$ mesons have been studied extensively in the standard model and its various extensions, for instance, in Refs.  \cite{Cheng:2017pcq,Ivanov:2019nqd,Barranco:2014bva,Barranco:2013tba,Akeroyd:2009tn,Dobrescu:2008er,Akeroyd:2007eh,Fajfer:2006uy,Fajfer:2005ug,Fajfer:2004mv,Akeroyd:2003jb,Akeroyd:2002pi}.
In this work,  we will systematically study the $D\to P/V/S\ell^+\nu_\ell$ decays with the SU(3) flavor symmetry. We  will  firstly construct the amplitude relations   between different decay modes of $D\to P\ell^+\nu_\ell$, $D\to V\ell^+\nu_\ell$ or $D\to S\ell^+\nu_\ell$ decays  by the SU(3) flavor symmetry and   the SU(3) flavor breaking.
We use  the available data to extract the SU(3) flavor symmetry/breaking  amplitudes and the form factors,  and then predict the  not yet measured  modes for further tests in experiments.
The forward-backward asymmetries $A^\ell_{FB}$, the lepton-side convexity parameters $ C^\ell_F$, the longitudinal polarizations of the final
charged lepton  $P^\ell_L$, the transverse polarizations of the final
charged lepton $P^\ell_T$,  the lepton spin asymmetries $A_{\lambda}$ and  the longitudinal polarization fractions $F_L$ of the final vector mesons   with two ways of integration  have also been predicted in the $D\to P/V\ell^+\nu_\ell$ decays. In addition, the $q^2$ dependence of some  differential observables   for the $D\to P/V\ell^+\nu_\ell$ decays are shown in figures.

This paper will be organized as follows. In Sec. II,  the theoretical framework in this work is presented,
including the effective hamiltonian, the hadronic helicity amplitude relations, the observables  and  the form factors.
The numerical results of the $D\to P/V/S\ell^+\nu_\ell$ semileptonic decays will be given in Sec. III.
Finally, we give the
summary and conclusion in Sec. IV.

\section{Theoretical Frame}

\subsection{The effective Hamiltonian}
In the standard model, the four-fermion charged-current effective Hamiltonian below the electroweak scale for the decays $D\rightarrow M\ell^+\nu_\ell~(M=P,V,S)$ can be written as
\begin{eqnarray}
\mathcal{H}_{eff}(c\rightarrow q\ell^+\nu_\ell)&=&\frac{G_F}{\sqrt{2}}V^*_{cq}\bar{q}\gamma^\mu(1-\gamma_5)c~\bar{\nu_\ell}\gamma_\mu(1-\gamma_5)\ell,\label{Heff}
\end{eqnarray}
with $q=s,d$.

The helicity amplitudes of the decays $D\rightarrow M\ell^+\nu_\ell$ can be written as
\begin{eqnarray}
\mathcal{M}(D\rightarrow M\ell^+\nu_\ell)&=&\frac{G_F}{\sqrt{2}}V_{cb}\sum_{mm'}g_{mm'} L^{\lambda_\ell\lambda_\nu}_mH^{\lambda_M}_{m'},
\end{eqnarray}
with
\begin{eqnarray}
L^{\lambda_\ell\lambda_\nu}_m&=&\epsilon_{\alpha}(m)\bar{\nu_\ell}\gamma^{\alpha}(1-\gamma_5)\ell,\\
H^{\lambda_M}_{m'}&=&\left\{\begin{array}{l}\epsilon{^*_{\beta}}(m')\langle{P/S}(p_{P/S})|\bar{q}\gamma^\beta(1-\gamma_5)c|D(p_D)\rangle\\
\epsilon^*_{\beta}(m')\langle{V}(p_V,\epsilon^*)|\bar{q}\gamma^\beta(1-\gamma_5)c|D(p_D)\rangle\end{array}\right.,\
\end{eqnarray}
where  the particle helicities $\lambda_M=0$ for $M=P/S$, $\lambda_M=0,\pm1$ for $M=V, \lambda_\ell=\pm\frac{1}{2}$ and $\lambda_\nu=+\frac{1}{2}$, as well as $\epsilon_{\mu}(m)$  is the polarization vectors of the virtual $W$ with $m=0,t,\pm1$.

The form factors of the  $D\to P$,  $D\to S$ and  $D\to V$ transitions are given by  \cite{Melikhov:2000yu,Cheng:2017pcq,Cheng:2003sm}
\begin{eqnarray}
\left<P(p)\left|\bar{d}_k\gamma_{\mu}c\right|D(p_D)\right>&=&f^P_+(q^2)(p+p_D)_{\mu}+\left[f^P_0(q^2)-f^P_+(q^2)\right]\frac{m^2_D-m^2_P}{q^2}q_{\mu},\\
\left<S(p)\left|\bar{d}_k\gamma_{\mu}\gamma_5c\right|D(p_D)\right>&=&-i\Big(f^S_+(q^2)(p+p_D)_{\mu}+\left[f^S_0(q^2)-f^S_+(q^2)\right]\frac{m^2_D-m^2_S}{q^2}q_{\mu}\Big),\\
\left<V(p,\varepsilon^*)\left|\bar{d}_k\gamma_{\mu}(1-\gamma_5)c\right|D(p_D)\right>
&=&\frac{2V^V(q^2)}{m_D+m_V}\epsilon_{\mu\nu\alpha\beta}\varepsilon^{*\nu}p^\alpha_Dp^\beta\nonumber\\
&&-i\left[\varepsilon^*_\mu(m_D+m_V)A^V_1(q^2)-(p_D+p)_\mu(\varepsilon^*.p_D)\frac{A^V_2(q^2)}{m_D+m_V}\right]\nonumber\\
&&+iq_\mu(\varepsilon^*.p_D)\frac{2m_V}{q^2}[A^V_3(q^2)-A^V_0(q^2)],
\end{eqnarray}
where $s=q^2$ ($q=p_D-p_M$), and $\varepsilon^*$ is the polarization of vector meson.
The hadronic helicity amplitudes can be written as
\begin{eqnarray}
H_\pm&=&0,\\
H_0&=&\frac{2m_{D_q}|\vec{p}_{P}|}{\sqrt{q^2}}f^{P}_+(q^2),\\
H_t&=&\frac{m_{D_q}^2-m_{P}^2}{\sqrt{q^2}}f^{P}_0(q^2),
\end{eqnarray}
for $D\to P\ell^+\nu_\ell$ decays,
\begin{eqnarray}
H_\pm&=&0,\\
H_0&=&\frac{i2m_{D_q}|\vec{p}_{S}|}{\sqrt{q^2}}f^{S}_+(q^2),\\
H_t&=&\frac{i m_{D_q}^2-m_{S}^2}{\sqrt{q^2}}f^{S}_0(q^2),
\end{eqnarray}
for $D\to S\ell^+\nu_\ell$ decays, and
\begin{eqnarray}
H_{\pm} &=&(m_{D_q}+m_V)A_1(q^2)\mp\frac{2m_{D_q}|\vec{p}_V|}{(m_{D_q}+m_V)}V(q^2), \\
H_{0}&=& \frac{1}{2m_V\sqrt{q^2}}\left[(m_{D_q}^2-m_V^2-q^2)(m_{D_q}+m_V)A_1(q^2)-\frac{4m_{D_q}^2|\vec{p}_V|^2}{m_{D_q}+m_V}A_2(q^2)\right], \\
H_{t}&=& \frac{2m_{D_q}|\vec{p}_V|}{\sqrt{q^2}}A_0(q^2),
\end{eqnarray}
for $D\to V\ell^+\nu_\ell$ decays, where $|\vec{p}_M|\equiv\sqrt{\lambda(m_{D_q}^2,m_M^2,q^2)}/2m_{D_q}$ with $\lambda(a,b,c)=a^2+b^2+c^2-2ab-2ac-2bc$.

\subsection{Hadronic helicity amplitude relations by the SU(3) flavor symmetry}

Charmed mesons containing one heavy $c$ quark are flavor SU(3) anti-triplets
\begin{eqnarray}
D_i=\Big( D^0(c\bar{u}),~D^+(c\bar{d}),~D^+_s(c\bar{s})\Big).
\end{eqnarray}
Light pseudoscalar $P$ and vector $V$ meson  octets and singlets  under the $SU(3)$ flavor symmetry of $u,d,s$ quarks are \cite{He:2018joe}
\begin{eqnarray}
 P&=&\left(\begin{array}{ccc}
\frac{\pi^0}{\sqrt{2}}+\frac{\eta_8}{\sqrt{6}}+\frac{\eta_1}{\sqrt{3}} & \pi^+ & K^+ \\
\pi^- &-\frac{\pi^0}{\sqrt{2}}+\frac{\eta_8}{\sqrt{6}}+\frac{\eta_1}{\sqrt{3}}  & K^0 \\
K^- & \overline{K}^0 &-\frac{2\eta_8}{\sqrt{6}}+\frac{\eta_1}{\sqrt{3}}
\end{array}\right)\,,\\
V&=&\left(\begin{array}{ccc}
\frac{\rho^0}{\sqrt{2}}+\frac{\omega_8}{\sqrt{6}}+\frac{\omega_1}{\sqrt{3}} & \rho^+ & K^{*+} \\
\rho^- &-\frac{\rho^0}{\sqrt{2}}+\frac{\omega_8}{\sqrt{6}}+\frac{\omega_1}{\sqrt{3}} & K^{*0} \\
K^{*-} & \overline{K}^{*0} &-\frac{2\omega_8}{\sqrt{6}}+\frac{\omega_1}{\sqrt{3}}
\end{array}\right)\,,
\end{eqnarray}
where $\omega$ and $\phi$ mix in an ideal form, and the $\eta$ and $\eta'$ ( $\omega$ and $\phi$) are mixtures of $\eta_1(\omega_1)=\frac{u\bar{u}+d\bar{d}+s\bar{s}}{\sqrt{3}}$ and $\eta_8(\omega_8)=\frac{u\bar{u}+d\bar{d}-2s\bar{s}}{\sqrt{6}}$ with the mixing angle $\theta_P$ ($\theta_V$). $\eta$ and $\eta'$ ($\omega$ and $\phi$)  are given by
\begin{eqnarray}
\left(\begin{array}{c}
\eta\\
\eta'
\end{array}\right)\,
=
\left(\begin{array}{cc}
cos\theta_P&-sin\theta_P\\
sin\theta_P&cos\theta_P
\end{array}\right)\,
\left(\begin{array}{c}
\eta_8\\
\eta_1
\end{array}\right),~~~~~~~~~
\left(\begin{array}{c}
\phi\\
\omega
\end{array}\right)\,
=
\left(\begin{array}{cc}
cos\theta_V&-sin\theta_V\\
sin\theta_V&cos\theta_V
\end{array}\right)\,
\left(\begin{array}{c}
\omega_8\\
\omega_1
\end{array}\right),
\end{eqnarray}\label{Eq:etamix}
where $\theta_P=[-20^\circ,-10^\circ]$ and $\theta_V=36.4^\circ$ from  Particle Data Group (PDG) \cite{PDG2022} will be used in our numerical  analysis.

The structures of the light scalar mesons  are   not fully understood
yet. Many suggestions are discussed, such as ordinary two quark states, four quark states, meson-meson bound states, molecular states,  glueball states or hybrid states, for examples, in Refs. \cite{Dai:2018fmx,Maiani:2004uc,tHooft:2008rus,Pelaez:2003dy,Sun:2010nv,Oller:1997ti,Baru:2003qq,Cheng:2005nb,Achasov:1996ei}.
In this work, we will consider the two quark and the four quark scenarios for the scalar mesons below or near 1 $GeV$.
In the two quark picture, the light scalar mesons can be written as   \cite{Momeni:2022gqb}
\begin{eqnarray}
S&=&\left(\begin{array}{ccc}
\frac{a^0_0}{\sqrt{2}}+\frac{\sigma}{\sqrt{2}} & a^+_0 & K^{+}_0 \\
a_0^- &-\frac{a_0^0}{\sqrt{2}}+\frac{\sigma}{\sqrt{2}} & K^{0}_0 \\
K^{-}_0 & \overline{K}^{0}_0 &f_0
\end{array}\right)\,.
\end{eqnarray}
The two isoscalars $f_0(980)$ and $f_0(500)$ are obtained by the mixing of $\sigma=\frac{u\bar{u}+d\bar{d}}{\sqrt{2}}$ and $f_0=s\bar{s}$
\begin{eqnarray}
\left(\begin{array}{c}
f_0(980)\\
f_0(500)
\end{array}\right)\,
=
\left(\begin{array}{cc}
\mbox{cos}\theta_S&\mbox{sin}\theta_S\\
-\mbox{sin}\theta_S&\mbox{cos}\theta_S
\end{array}\right)\,\left(\begin{array}{c}
f_0\\
\sigma
\end{array}\right)\,,
\end{eqnarray}\label{Eq:f0mix2q}
where the three possible ranges of the mixing angle, $25^\circ<\theta_S<40^\circ$, $140^\circ<\theta_S<165^\circ $ and $~-30^\circ<\theta_S<30^\circ $ \cite{Cheng:2005nb,LHCb:2013dkk} will be analyzed in our numerical results.
In the four quark picture, the light scalar mesons  are given as   \cite{Jaffe:1976ig,PDG2022}
\begin{eqnarray}
&&\sigma=u\bar{u}d\bar{d},~~~~~~~~~~~~~~~~~~~~~~~~~~f_0=(u\bar{u}+d\bar{d})s\bar{s}/\sqrt{2},\nonumber\\
&&a^0_0=(u\bar{u}-d\bar{d})s\bar{s}/\sqrt{2},~~~~~~~~~~~~~~~a^+_0=u\bar{d}s\bar{s},~~~~~~~~~~~~~~~~a^-_0=d\bar{u}s\bar{s},\nonumber\\
&& K^+_0=u\bar{s}d\bar{d},~~~~~~~~~K^0_0=d\bar{s}u\bar{u},~~~~~~~~~\bar{K}^0_0=s\bar{d}u\bar{u},~~~~~~~~~K^+_0=s\bar{u}d\bar{d},
\end{eqnarray}
and the two isoscalars are expressed as
\begin{eqnarray}
\left(\begin{array}{c}
f_0(980)\\
f_0(500)
\end{array}\right)\,
=
\left(\begin{array}{cc}
\mbox{cos}\phi_S&\mbox{sin}\phi_S\\
-\mbox{sin}\phi_S&\mbox{cos}\phi_S
\end{array}\right)\,\left(\begin{array}{c}
f_0\\
\sigma
\end{array}\right)\,,
\end{eqnarray}\label{Eq:f0mix4q}
where the constrained mixing angle $\phi_S=(174.6^{+3.4}_{-3.2})^\circ$ \cite{Maiani:2004uc}.

In terms  of the  SU(3) flavor symmetry, meson states and quark operators can be parameterized into SU(3) tensor forms, while the leptonic helicity amplitudes $L^{\lambda_\ell,\lambda_\nu}_{m}$ are invariant under the SU(3)  flavor symmetry. And   the hadronic helicity amplitude relations of the  $D\rightarrow M\ell^+\nu_\ell(M=P,V,S)$ decays can be parameterized as
\begin{eqnarray}
H(D\rightarrow M\ell^+\nu_\ell)=c_{0}^M D_iM^i_jH^j, \label{Eq:HM}
\end{eqnarray}
where $H^2\equiv V^*_{cd}$ and $H^3\equiv V^*_{cs}$ are the CKM matrix elements, and $c_{0}^M$ are the nonperturbative
coefficients of the  $D\rightarrow M\ell^+\nu_\ell$ decays under the SU(3) flavor symmetry.  Noted that the hadronic helicity amplitudes for the $D\rightarrow S\ell^+\nu_\ell$ decays in Eq. (\ref{Eq:HM}) are given in the two quark picture of the light scalar mesons, and ones in the four quark picture of the light scalar mesons will be given later.

The SU(3) flavor breaking effects  mainly come from different masses of $u$, $d$ and $s$ quarks. Following Ref. \cite{Xu:2013dta}, the SU(3) breaking amplitudes of the $D\rightarrow M\ell^+\nu_\ell$ decays can be give as
\begin{eqnarray}
\Delta H(D\rightarrow M\ell^+\nu_\ell)=c_1^M D_aW^a_iM^i_jH^j+c_2^M D_iM^i_aW^a_jH^j,  \label{Eq:D2MlvSU3B}
\end{eqnarray}
with
\begin{eqnarray}
 W=\big(W^i_j\big)=\left(\begin{array}{ccc}
1 & 0&0\\
0 &1 & 0 \\
0 & 0 &-2
\end{array}\right),
\end{eqnarray}
where $c_{1,2}^M$ are the nonperturbative    SU(3) flavor  breaking
coefficients.

In the four quark picture of the light scalar mesons,   the hadronic helicity amplitudes of the  $D\rightarrow S\ell^+\nu_\ell $ decays under  the SU(3) flavor symmetry   are
\begin{eqnarray}
H(D\rightarrow S\ell^+\nu_\ell)^{4q}=c'^S_0 D_iS^{im}_{jm}H^j. \label{Eq:HS4q}
\end{eqnarray}
And the corresponding  SU(3) flavor breaking amplitudes of the $D\rightarrow S\ell^+\nu_\ell$ decays are
\begin{eqnarray}
\Delta H(D\rightarrow S\ell^+\nu_\ell)^{4q}=c'^S_1 D_aW^a_iS^{im}_{jm}H^j+c'^S_2 D_iS^{im}_{am}W^a_jH^j+c'^S_1 D_iS^{im}_{ja}W^a_mH^j. \label{Eq:HS4qSU3B}
\end{eqnarray}

 In terms  of the  SU(3) flavor symmetry, the hadronic helicity amplitude relations  for the $D\rightarrow P\ell^+\nu_\ell $,  $D\rightarrow V\ell^+\nu_\ell$ and $D\rightarrow S\ell^+\nu_\ell$ decays are   summarized in later Tab. \ref{Tab:HD2PlvAmp}, Tab. \ref{Tab:HD2VlvAmp} and Tab. \ref{Tab:HD2SlvAmp},  respectively.

\subsection{Observables for the $D\to M \ell^+\nu_\ell$  decays }

The double differential branching ratios  of the  $D\to M \ell^+\nu_\ell$ decays are \cite{Ivanov:2019nqd}
\begin{eqnarray}
  \frac{d\mathcal{B}(D\to
  M\ell^+\nu_\ell)}{dq^2d(\cos\theta)}&=&\frac{\tau_{D}G_F^2|V_{cq}|^2\lambda^{1/2}(q^2-m_\ell^2)^2}{64(2\pi)^3M_{D_{(s)}}^3q^2}
  \Biggl[(1+\cos^2\theta)
{\cal H}_U+2\sin^2\theta{\cal H}_L+2\cos\theta{\cal H}_P\cr
&&+\frac{m_\ell^2}{q^2}(\sin^2\theta{\cal H}_U
   +2\cos^2\theta{\cal H}_L+ 2 {\cal H}_S-4\cos\theta {\cal H}_{SL})\Bigr],  \label{eq:dGamma}
\end{eqnarray}
where  $\lambda\equiv
\lambda(m_{D_q}^2,m_M^2,q^2)$, $m_\ell^2\leq q^2\leq(m_{D_q}-m_M)^2$, and
\begin{equation}
{\cal H}_U=|H_+|^2+|H_-|^2, \quad {\cal H}_L=|H_0|^2, \quad {\cal
   H}_P=|H_+|^2-|H_-|^2,\quad
    {\cal H}_S=|H_t|^2, \quad {\cal H}_{SL}=\Re(H_0H_t^\dag).  \label{eq:hh}
  \end{equation}

The differential branching ratios  integrated
over   $\cos\theta$ are  \cite{Ivanov:2019nqd}
\begin{equation}
 \frac{d\mathcal{B}(D_{(s)}\to
  M\ell^+\nu_\ell)}{dq^2}=\frac{\tau_{D}G_F^2 |V_{cq}|^2\lambda^{1/2}(q^2-m_\ell^2)^2}{24(2\pi)^3M_{D_{(s)}}^3q^2}
 {\cal H}_{\rm total},\label{Eq:dbdq2}
\end{equation}
with
\begin{equation}
  \label{eq:htot}
 {\cal H}_{\rm total}\equiv ({\cal H}_U+{\cal
   H}_L)\left(1+\frac{m_\ell^2}{2q^2}\right) +\frac{3m_\ell^2}{2q^2}{\cal H}_S.
\end{equation}

The lepton flavor universality in $D_{(s)}\to
  M\ell^+\nu_\ell$  is defined in
a manner identical $R^{\mu/e}$ as
\begin{eqnarray}
R^{\mu/e}=\frac{\int^{q_{max}}_{q_{min}}d\mathcal{B}(D_{(s)}\to
  M\mu^+\nu_\mu)/dq^2}{\int^{q_{max}}_{q_{min}}d\mathcal{B}(D_{(s)}\to
  Me^+\nu_e)/dq^2}.
\end{eqnarray}

The forward-backward asymmetries  are defined as \cite{Ivanov:2019nqd}
\begin{eqnarray}
A_{FB}^\ell(q^2)&=&\frac{\int^0_{-1}d\mbox{cos}\theta_\ell~ \frac{d\mathcal{B}(D\to M\ell\nu)}{dq^2d\mbox{cos}\theta_\ell}-\int^1_{0}d\mbox{cos}\theta_\ell \frac{d\mathcal{B}(D\to M\ell\nu)}{dq^2d\mbox{cos}\theta_\ell}}{\int^0_{-1}d\mbox{cos}\theta_\ell~ \frac{d\mathcal{B}(D\to M\ell\nu)}{dq^2d\mbox{cos}\theta_\ell}+\int^1_{0}d\mbox{cos}\theta_\ell \frac{d\mathcal{B}(D\to M\ell\nu)}{dq^2d\mbox{cos}\theta_\ell}}\\
&=&\frac34\frac{{\cal
      H}_P-\frac{2m_\ell^2}{q^2}{\cal H}_{SL}}{{\cal H}_{\rm total}}.
\end{eqnarray}

The lepton-side convexity parameters are given by \cite{Ivanov:2019nqd}
\begin{equation}
C^\ell_F(q^2)=\frac34\left(1-\frac{m_\ell^2}{q^2}\right)\frac{{\cal H}_U-2{\cal H}_L}{{\cal H}_{\rm total}}.   \label{eq:clf}
\end{equation}

The longitudinal polarizations of the final
charged lepton $\ell$ are defined by  \cite{Ivanov:2019nqd}
\begin{equation}
P_L^\ell(q^2)=\frac{({\cal H}_U+{\cal
   H}_L)\left(1-\frac{m_\ell^2}{2q^2}\right) -\frac{3m_\ell^2}{2q^2}{\cal H}_S}{{\cal H}_{\rm total}},   \label{eq:ple}
\end{equation}
and its transverse polarizations are
\begin{equation}
  P_T^\ell(q^2)=-\frac{3\pi m_\ell}{8\sqrt{q^2}}\frac{{\cal H}_P+2{\cal H}_{SL}}{{\cal H}_{\rm total}}.   \label{eq:pte}
\end{equation}

The lepton spin asymmetry in the $\ell-\bar{\nu}_\ell$ center of mass frame is defined by \cite{Fajfer:2012vx,Tanaka:1994ay,Celis:2012dk,Tanaka:2010se}
\begin{eqnarray}
A_{\lambda}(q^2)&=&\frac{d\mathcal{B}(D\to M\ell^+\nu_\ell)[\lambda_\ell=-\frac{1}{2}]/dq^2-d\mathcal{B}(D\to M\ell^+\nu_\ell)[\lambda_\ell=+\frac{1}{2}]/dq^2}
{d\mathcal{B}(D\to M\ell^+\nu_\ell)[\lambda_\ell=-\frac{1}{2}]/dq^2+d\mathcal{B}(D\to M\ell^+\nu_\ell)[\lambda_\ell=+\frac{1}{2}]/dq^2}
\\
&=&\frac{{\cal H}_{\rm total}-\frac{6m_\ell^2}{2q^2}{\cal H}_{\rm S}}{{\cal H}_{\rm total}}.
\end{eqnarray}

For the $D\to V\ell^+\nu_\ell$ decays, the longitudinal polarization fractions
of the final vector mesons are given by \cite{Ivanov:2019nqd}
\begin{equation}
 F_L(q^2)=\frac{{\cal
   H}_L\left(1+\frac{m_\ell^2}{2q^2}\right) +\frac{3m_\ell^2}{2q^2}{\cal H}_S}{{\cal H}_{\rm total}},  \label{eq:fl}
\end{equation}
then its transverse polarization fraction $F_T(q^2)=1- F_L(q^2)$.

Noted that, for  $q^2$-integration of $X(q^2)= A^\ell_{FB},~C^\ell_F,~P^\ell_L,~P^\ell_T,~A_{\lambda}$ and $F_L$, following Ref. \cite{Bobeth:2010wg}, two ways of integration are considered.
The normalized $q^2$-integrated observables $\langle X \rangle$  are calculated by separately
integrating the numerators and denominators with the same $q^2$ bins.
The ``naively integrated" observables are obtained by
\begin{eqnarray}
\overline{X} =\frac{1}{q^2_{max}-q^2_{min}}\int^{q^2_{max}}_{q^2_{min}}dq^2X(q^2).
\end{eqnarray}

\subsection{Form factors}
In order to obtain more precise  observables, one also need  considering  the $q^2$ dependence of the form factors for  the  $D\to P\ell^+\nu_\ell$, $D\to V\ell^+\nu_\ell$  and $D\to S\ell^+\nu_\ell$ decays.
The following  cases will be considered in our analysis of $D\to P/V\ell^+\nu_\ell$ decays.
\begin{itemize}
\item[\bf$C_1$:] All form factors are treated as constants
without the hadronic momentum-transfer $q^2$  dependence, and different form factors are related by the SU(3) flavor symmetry, $i.e.$, the SU(3) flavor breaking terms such as $c^M_{1,2}$ and $c'^S_{1,2,3}$ in later  Tabs. \ref{Tab:HD2PlvAmp}, \ref{Tab:HD2VlvAmp} and  \ref{Tab:HD2SlvAmp} are ignored.

\item[\bf $C_2$:] With the SU(3) flavor symmetry, the modified pole model for
the $q^2$-dependence of $F_i(q^2)$ is used \cite{Li:2020ylu}
 \begin{eqnarray}
F_i(q^2)=\frac{F_i(0)}{\left(1-\frac{q^2}{m^2_{pole}}\right)\left(1-\alpha_i\frac{q^4}{m^4_{pole}}\right)},
\end{eqnarray}
where $m_{pole}=m_{D^{*+}}$ for $c\to d \ell^+\nu_\ell$ transitions and $m_{pole}=m_{D^{*+}_s}$ for $c\to s \ell^+\nu_\ell$ transitions, and $\alpha_i$ are free parameters and are different for $f^P_+(q^2)$, $f^P_0(q^2)$, $V(q^2)$, $A_1(q^2)$ and $A_2(q^2)$, we will take $\alpha_i\in[-1,1]$ in our analysis.

\item[\bf $C_3$:]  With the SU(3) flavor symmetry, following Ref. \cite{Melikhov:2000yu}
 \begin{eqnarray}
F_i(q^2)&=&\frac{F_i(0)}{\left(1-\frac{q^2}{m^2_{pole}}\right)\left(1-\sigma_{1i}\frac{q^2}{m^2_{pole}}+\sigma_{2i}\frac{q^4}{m^4_{pole}}\right)}  ~~~~~~~~~~\mbox{for $f^P_+(q^2)$ and $V(q^2)$},\\
F_i(q^2)&=&\frac{F_i(0)}{\left(1-\sigma_{1i}\frac{q^2}{m^2_{pole}}+\sigma_{2i}\frac{q^4}{m^4_{pole}}\right)}~~~~~~~~~~ \mbox{for $f^P_0(q^2)$,  $A_1(q^2)$ and $A_2(q^2)$},
\end{eqnarray}
where $\sigma_{1,2}$ for the $D\to \pi$ and $D\to K^*$ transitions    from Ref. \cite{Melikhov:2000yu} will be used in our results.

\item[\bf $C_4$:]  Considering  the SU(3) flavor breaking terms such as $c^M_{1,2}$   and $c'^S_{1,2,3}$ in later Tabs. \ref{Tab:HD2PlvAmp}, \ref{Tab:HD2VlvAmp} and \ref{Tab:HD2SlvAmp},  the form factors in $C_3$ case are used.
\end{itemize}

As for the form factors of the $D\to S\ell^+\nu_\ell$ decays, we find that the vector dominance model \cite{Achasov:2012kk} and  the double pole model \cite{Soni:2020sgn}  give the similar SU(3) flavor symmetry predictions for the branching ratios of the $D\to S\ell^+\nu_\ell$ decays.  The following form factors from  the  vector dominance model will be used in  the numerical results,
 \begin{eqnarray}
F_i(q^2)=\frac{F_i(0)}{\left(1-q^2/m^2_{pole}\right)}  ~~~~~~~~~\mbox{ for $f^S_{+}(q^2)$  and   $f^S_{0}(q^2)$}.
\end{eqnarray}

After considering above $q^2$ dependence, we only need to focus on the $F_i(0)$.  Since these form factors  $F_i(0)$ also preserve the SU(3) flavor symmetry, the same relations in Tabs. \ref{Tab:HD2PlvAmp}, \ref{Tab:HD2VlvAmp} and \ref{Tab:HD2SlvAmp} will be used for $F_i(0)$.
If considering  the form factors ratios $f_+(0)/f_0(0)=1$ for  $D\to P/S\ell^+\nu_\ell$ decays, $r_V\equiv V(0)/A_1(0)=1.46\pm0.07$,  $r_2\equiv A_2(0)/A_1(0)=0.68\pm0.06$ in $D^0\to K^{*-}\ell^+\nu_\ell$ decays from PDG \cite{PDG2022} and  the SU(3) flavor symmetry,   there is only one free form factor  $f^{P,S}_+(0)$ and $A_1(0)$ for the $D\to P/S\ell^+\nu_\ell$  and $D\to V \ell^+\nu_\ell$ decays, respectively. As a result, the branching ratios only depend on one form factor $f^P_+(0)$, $f^S_+(0)$ or $A_1(0)$ and  the CKM matrix element $V_{cq}$.

\section{Numerical results}

The theoretical input parameters and the experimental
data within the $2\sigma$ errors from PDG  \cite{PDG2022} will
be used in our numerical results.

\subsection{$D\to P\ell^+\nu_\ell$ decays}

Considering  both the  SU(3) flavor symmetry and the SU(3) flavor breaking contributions, the hadronic helicity amplitudes  for the $D\rightarrow P\ell^+\nu_\ell$  decays are   given  in  Tab. \ref{Tab:HD2PlvAmp},
in which   we keep the CKM matrix element $V_{cs}$ and $V_{cd}$ information for comparing conveniently.  In addition,  $H(D^+_s\to \pi^0\ell^+\nu_\ell)$ are obtained  by  neutral meson mixing with $\delta^2=(5.18\pm0.71)\times10^{-4}$ in Ref. \cite{Li:2020ylu}.    From Tab. \ref{Tab:HD2PlvAmp},   we can easily see the hadronic helicity amplitude relations of  the  $D\rightarrow P\ell^+\nu_\ell$  decays.
There are four  nonperturbative  parameters $A_{1,2,3,4}$ in the  $D\rightarrow P\ell^+\nu_\ell$ decays with $A_1\equiv c^P_0+c^P_1-2c^P_2$, $A_2\equiv c^P_0-2c^P_1-2c^P_2$, $A_3\equiv c^P_0+c^P_1+c^P_2$ and $A_4\equiv c^P_0-2c^P_1+c^P_2$.  If  neglecting the SU(3) flavor breaking $c^P_1$ and $c^P_2$ terms, $A_1=A_2=A_3=A_4=c^P_0$, and then all hadronic helicity amplitudes are related by only one parameter $c^P_0$.
\begin{table}[b]
\renewcommand\arraystretch{1.2}
\tabcolsep 0.1in
\centering
\caption{The hadronic helicity amplitudes for the $D\to P\ell^+\nu$ decays including  both the SU(3) flavor symmetry and the SU(3) flavor breaking contributions.  $A_1\equiv c^P_0+c^P_1-2c^P_2$, $A_2\equiv c^P_0-2c^P_1-2c^P_2$, $A_3\equiv c^P_0+c^P_1+c^P_2$, $A_4\equiv c^P_0-2c^P_1+c^P_2$.  $A_1=A_2=A_3=A_4=c^P_0$ if  neglecting the SU(3) flavor breaking $c^P_1$ and $c^P_2$ terms.             }\vspace{0.1cm}
{\footnotesize
\begin{tabular}{lccccc}  \hline\hline
Hadronic helicity amplitudes  & SU(3) flavor amplitudes\\\hline
$H(D^0\to K^-\ell^+\nu_\ell)$&$A_1V^*_{cs}$\\
$H(D^+\to \overline{K}^0\ell^+\nu_\ell)$&$A_1V^*_{cs}$\\
%
%
%
$H(D^+_s\to \eta\ell^+\nu_\ell)$&$\big(-cos\theta_P\sqrt{2/3}-sin\theta_P/\sqrt{3}\big)A_2V^*_{cs}$ \\
$H(D^+_s\to \eta'\ell^+\nu_\ell)$&$\big(-sin\theta_P\sqrt{2/3}+cos\theta_P/\sqrt{3}\big)A_2V^*_{cs}$\\
$H(D^+_s\to \pi^0\ell^+\nu_\ell)$&$-\delta \big(-cos\theta_P\sqrt{2/3}-sin\theta_P/\sqrt{3}\big)A_2V^*_{cs}$ \\\hline
$H(D^0\to \pi^-\ell^+\nu_\ell)$&$A_3V^*_{cd}$\\
$H(D^+\to \pi^0\ell^+\nu_\ell)$&$-\frac{1}{\sqrt{2}}A_3V^*_{cd}$ \\
%
%
%
$H(D^+\to \eta\ell^+\nu_\ell)$&$\big(cos\theta_P/\sqrt{6}-sin\theta_P/\sqrt{3}\big)A_3V^*_{cd}$ \\
$H(D^+\to \eta'\ell^+\nu_\ell)$&$\big(sin\theta_P/\sqrt{6}+cos\theta_P/\sqrt{3}\big)A_3V^*_{cd}$ \\
$H(D^+_s\to K^0\ell^+\nu_\ell)$&$A_4V^*_{cd}$  \\\hline
\end{tabular}\label{Tab:HD2PlvAmp}}
\end{table}

Many decay modes of the $D\rightarrow Pe^+\nu_e,P\mu^+\nu_\mu$ decays have been measured, and the experimental data with $2\sigma$ errors are listed in the second column of Tab. \ref{Tab:BrD2Plv}.
One can constrain the  parameters $A_i$  by the present experimental data within $2\sigma$  errors and then predict other not yet measured  branching ratios. Four cases $C_{1,2,3,4}$ will be considered
in our analysis.  The numerical results of $\mathcal{B}(D\to P\ell^+\nu_\ell)$ in the $C_{1}$, $C_{2}$, $C_{3}$ and $C_{4}$ cases  are given in the third, forth, fifth and sixth columns of Tab. \ref{Tab:BrD2Plv}, respectively. And our comments on the results are as follows.
\begin{table}[b]
\renewcommand\arraystretch{1.3}
\tabcolsep 0.1in
\centering
\caption{Branching ratios of the $D\to P\ell^+\nu$ decays.  $^\dag$Denotes that the corresponding experimental data from PDG \cite{PDG2022} are not used to constrain $A_i$ in this case.  }\vspace{0.1cm}
{\footnotesize
\begin{tabular}{lcccccc}  \hline\hline
Branching ratios                                             &    Exp. data        &    Ones in  $C_1$    &    Ones in $C_2$      &    Ones in $C_3$       &    Ones in $C_4$    &    Previous ones  \\\hline
$\mathcal{B}(D^+\to \overline{K}^0e^+\nu_e)(\times10^{-2})$  &    $8.72\pm0.18$    &    $8.84\pm0.06$    &    $8.83\pm0.07$       &    $8.84\pm0.06$       &    $8.83\pm0.07$    &    $$ \\
$\mathcal{B}(D^+\to \pi^0e^+\nu_e)(\times10^{-3})$           &    $3.72\pm0.34$    &    $3.75\pm0.05$    &    $5.40\pm1.33^\dag$ &    $5.04\pm0.12^\dag$ &    $3.70\pm0.11$    &    $$ \\
$\mathcal{B}(D^+\to \eta e^+\nu_e)(\times10^{-3})$           &    $1.11\pm0.14$    &    $1.15\pm0.05$    &    $1.20\pm0.05$       &    $1.20\pm0.05$       &    $0.92\pm0.08$    &    $$ \\
$\mathcal{B}(D^+\to \eta'e^+\nu_e)(\times10^{-4})$           &    $2.0\pm0.8$      &    $2.59\pm0.14$    &    $2.22\pm0.34$       &    $2.09\pm0.14$       &   $1.50\pm0.20$     &    $$ \\
$\mathcal{B}(D^0\to K^-e^+\nu_e)(\times10^{-2})$             &  $3.549\pm0.052$    &    $3.52\pm0.02$    &    $3.52\pm0.03$       &    $3.52\pm0.03$       &    $3.52\pm0.02$    &    $$ \\
$\mathcal{B}(D^0\to \pi^-e^+\nu_e)(\times10^{-3})$           &    $2.91\pm0.08$    &    $2.95\pm0.03$    &    $4.23\pm1.03^\dag$ &    $3.97\pm0.09^\dag$ &    $2.89\pm0.06$    &    $$ \\
$\mathcal{B}(D^+_s\to \eta e^+\nu_e)(\times10^{-2})$         &    $2.32\pm0.16$    &    $2.37\pm0.11$    &    $2.34\pm0.14$       &    $2.36\pm0.12$       &    $2.32\pm0.16$    &    $$ \\
$\mathcal{B}(D^+_s\to \eta'e^+\nu_e)(\times10^{-3})$         &     $8.0\pm1.4$     &    $9.05\pm0.04$    &    $8.25\pm1.13$       &    $8.04\pm0.43$       &    $8.02\pm1.38$    &    $$ \\
$\mathcal{B}(D^+_s\to K^0e^+\nu_e)(\times10^{-3})$           &     $3.4\pm0.8$     &    $3.10\pm0.08$    &    $3.56\pm0.39$       &    $3.54\pm0.12$       &    $3.40\pm0.80$    &    $$ \\
$\mathcal{B}(D^+_s\to \pi^0e^+\nu_e)(\times10^{-5})$         &    $\cdots$         &    $1.51\pm0.07$    &    $2.10\pm0.56$       &    $1.96\pm0.10$       &    $1.92\pm0.13$    &    $2.65\pm0.38$~\cite{Li:2020ylu} \\\hline
$\mathcal{B}(D^+\to \overline{K}^0\mu^+\nu_\mu)(\times10^{-2})$  &  $8.76\pm0.38$  &    $8.56\pm0.06$    &    $8.69\pm0.15$       &    $8.61\pm0.06$       &    $8.61\pm0.06$    &    $$\\
$\mathcal{B}(D^+\to \pi^0\mu^+\nu_\mu)(\times10^{-3})$        &    $3.50\pm0.30$   &    $3.67\pm0.05$    &    $5.32\pm1.31^\dag$ &    $4.96\pm0.12^\dag$ &    $3.64\pm0.10$    &    $$\\
$\mathcal{B}(D^+\to \eta \mu^+\nu_\mu)(\times10^{-3})$        &   $1.04\pm0.22$    &    $1.11\pm0.05$    &    $1.18\pm0.07$       &    $1.17\pm0.05$       &    $0.90\pm0.08$    &    $^{1.21~\mbox{\cite{Faustov:2019mqr}}}_{0.75\pm0.15~\mbox{\cite{Leng:2020fei}}}$\\
$\mathcal{B}(D^+\to \eta'\mu^+\nu_\mu)(\times10^{-4})$        &    $\cdots$        &    $2.42\pm0.13$    &    $2.10\pm0.33$       &    $1.96\pm0.13$       &    $1.41\pm0.19$    &    $^{2.11~\mbox{\cite{Faustov:2019mqr}}}_{1.06\pm0.20~\mbox{\cite{Leng:2020fei}}}$\\
$\mathcal{B}(D^0\to K^-\mu^+\nu_\mu)(\times10^{-2})$          &    $3.41\pm0.08$   &    $3.41\pm0.02$    &    $3.44\pm0.05$       &    $3.43\pm0.02$       &    $3.43\pm0.02$    &    $$ \\
$\mathcal{B}(D^0\to \pi^-\mu^+\nu_\mu)(\times10^{-3})$        &    $2.67\pm0.24$   &    $2.89\pm0.02$    &    $4.17\pm1.01^\dag$ &    $3.90\pm0.09^\dag$ &    $2.85\pm0.06$    &    $$ \\
$\mathcal{B}(D^+_s\to \eta \mu^+\nu_\mu)(\times10^{-2})$      &    $2.4\pm1.0$     &    $2.30\pm0.10$    &    $2.30\pm0.17$       &    $2.31\pm0.12$       &    $2.26\pm0.16$    &    $$ \\
$\mathcal{B}(D^+_s\to \eta'\mu^+\nu_\mu)(\times10^{-2})$      &    $1.1\pm1.0$     &    $0.86\pm0.03$    &    $0.79\pm0.11$       &    $0.77\pm0.04$       &    $0.76\pm0.13$    &    $$ \\
$\mathcal{B}(D^+_s\to K^0\mu^+\nu_\mu)(\times10^{-3})$        &    $\cdots$        &    $3.01\pm0.08$    &    $3.51\pm0.38$       &    $3.46\pm0.11$       &    $3.33\pm0.78$    &    $^{3.9~\mbox{\cite{Faustov:2019mqr}}}_{3.85\pm0.76~\mbox{\cite{Leng:2020fei}}}$\\\
$\mathcal{B}(D^+_s\to \pi^0\mu^+\nu_\mu)(\times10^{-5})$      &    $\cdots$        &    $1.48\pm0.07$    &    $2.09\pm0.53$       &    $1.93\pm0.10$       &    $1.89\pm0.13$    &    $$ \\\hline
$\mathcal{B}(D^+_s\to \pi^0\tau^+\nu_\tau)(\times10^{-10})$        &    $\cdots$        &    $3.45\pm0.21$    &    $160.34\pm149.53$    &    $4.20\pm0.26$       &    $4.08\pm0.34$    &    $(27\sim36)~\mbox{\cite{Li:2020ylu}}$  \\
\hline
$R^{\mu/e}(D^+\to \overline{K}^0\ell^+\nu_\ell)$             &          $$          &      $0.969$      &        $0.984\pm0.013$        &      $0.974$      &      $0.974$    &    $$\\
$R^{\mu/e}(D^+\to \pi^0\ell^+\nu_\ell)$                      &          $$          &      $0.977$      &        $1.009\pm0.026$        &      $0.984$      &      $0.984$    &    $$\\
$R^{\mu/e}(D^+\to \eta \ell^+\nu_\ell)$                      &          $$          &      $0.967$      &        $0.984\pm0.014$        &      $0.973$      &      $0.973$    &    $$\\
$R^{\mu/e}(D^+\to \eta'\ell^+\nu_\ell)$                      &          $$          &      $0.935$      &        $0.948\pm0.012$        &      $0.940$      &      $0.940$    &    $$\\
$R^{\mu/e}(D^0\to K^-\ell^+\nu_\ell)$                        &          $$          &      $0.969$      &        $0.984\pm0.013$        &      $0.974$      &      $0.974$    &    $$\\
$R^{\mu/e}(D^0\to \pi^-\ell^+\nu_\ell)$                      &          $$          &      $0.977$      &        $1.008\pm0.026$        &      $0.984$      &      $0.984$    &    $$\\
$R^{\mu/e}(D^+_s\to \eta \ell^+\nu_\ell)$                    &          $$          &      $0.971$      &        $0.987\pm0.013$        &      $0.976$      &      $0.976$    &    $$\\
$R^{\mu/e}(D^+_s\to \eta'\ell^+\nu_\ell)$                    &          $$          &      $0.946$      &        $0.958\pm0.011$        &      $0.952$      &      $0.952$    &    $$\\
$R^{\mu/e}(D^+_s\to K^0\ell^+\nu_\ell)$                      &          $$          &      $0.973$      &        $0.992\pm0.016$        &      $0.978$      &      $0.978$    &    $$\\
$R^{\mu/e}(D^+_s\to \pi^0\ell^+\nu_\ell)$                    &          $$          &      $0.980$      &        $1.010\pm0.025$        &      $0.985$      &      $0.985$    &    $$\\
\hline
\end{tabular}\label{Tab:BrD2Plv}}
\end{table}
\begin{itemize}
\item {\bf Results in $C_1$ case: }
From the third column of Tab. \ref{Tab:BrD2Plv},  one can see that the SU(3) flavor symmetry predictions of  $\mathcal{B}(D\to P\ell^+\nu_\ell)$ in the $C_1$ case  are  entirely  consistent  with all present experiential data.
The not yet measured branching ratios of  the $D^+_s\to \pi^0e^+\nu_e$, $D^+_s\to \pi^0\mu^+\nu_\mu$,  $D^+\to \eta'\mu^+\nu_\mu$ and $D^+_s\to K^0\mu^+\nu_\mu$ decays  are predicted on the order of $\mathcal{O}(10^{-3}-10^{-5})$, nevertheless, $\mathcal{B}(D^+_s\to \pi^0\tau^+\nu_\tau)$ is  predicted on the order of $\mathcal{O}(10^{-10})$ due to its narrow phase space and $(q^2-m_\tau^2)^2$ suppression of
the differential branching ratios   in Eq. (\ref{Eq:dbdq2}).

\item {\bf Results in $C_{2,3}$ cases: }
The numerical results in $C_{2,3}$ cases are similar. The experimental upper limits of $\mathcal{B}(D^+\to \pi^0\ell^+\nu_\ell)$ and $\mathcal{B}(D^0\to \pi^-\ell^+\nu_\ell)$ have not been used to constrain  the predictions of  $\mathcal{B}(D\to P\ell^+\nu_\ell)$,  since the upper limits of the predictions of $\mathcal{B}(D^+\to \pi^0\ell^+\nu_\ell)$ and $\mathcal{B}(D^0\to \pi^-\ell^+\nu_\ell)$  by the SU(3) flavor symmetry  in $C_{2,3}$ cases are slightly larger than their experimental data.  Other SU(3) flavor symmetry predictions are consistent with their experimental data within $2\sigma$ errors.

\item {\bf Results in $C_{4}$ case: } As given in the sixth column of Tab. \ref{Tab:BrD2Plv},  if considering both the hadronic momentum-transfer $q^2$  dependence of the form factors and the SU(3) flavor breaking contributions,   all SU(3) flavor symmetry predictions are consistent with their experimental data within $2\sigma$ errors. For some decays, the errors of the theoretical predictions are much smaller than ones of their experimental data.

\item The previous predictions for the not yet measured  branching ratios are listed in the last column of Tab. \ref{Tab:BrD2Plv}, our predictions are in the same order of magnitude as previous ones for the  $D\to Pe^+\nu_e,P\mu^+\nu_\mu$ decays. And our prediction of  $\mathcal{B}(D^+_s\to \pi^0\tau^+\nu_\tau)$ is one order smaller than previous one in Ref. \cite{Li:2020ylu}.

\item In addition, the lepton flavor universality parameters $R^{\mu/e}(D\to P\ell^+\nu_\ell)$  are also given in Tab. \ref{Tab:BrD2Plv}, since many terms are canceled in the ratios, these predictions are quite accurate,  and all processes have similar results.

\end{itemize}

\begin{figure}[hb]
\begin{center}
\includegraphics[scale=0.9]{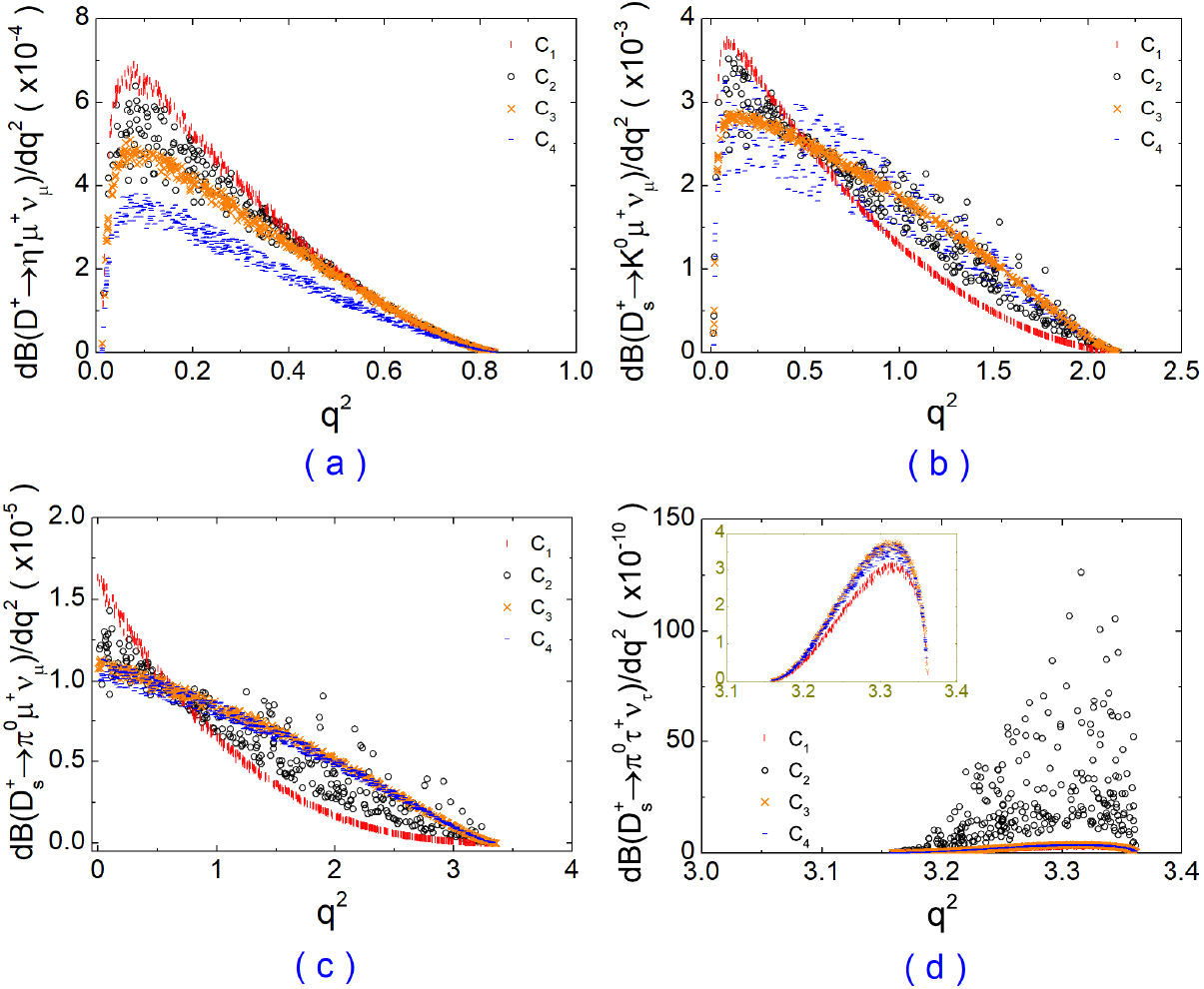}
\end{center}
\caption{The $q^2$ dependence of the differential branching ratios for some $D\to P\ell^+\nu_\ell$ with present experimental bounds. }\label{fig1dBP}
\end{figure}

For the $q^2$ dependence of the differential branching ratios of the $D\to P\ell^+\nu_\ell$ decays with present experimental bounds, we only show  the not yet measured  processes $D^+\to \eta'\mu^+\nu_\mu, ~D^+_s\to K^0\mu^+\nu_\mu,~D^+_s\to \pi^0\mu^+\nu_\mu$ and $D^+_s\to \pi^0\tau^+\nu_\tau$  in Fig. \ref{fig1dBP}.
 We do not show $d\mathcal{B}(D^+_s\to \pi^0e^+\nu_e)/dq^2$, since it is similar to $d\mathcal{B}(D^+_s\to \pi^0\mu^+\nu_\mu)/dq^2$ in Fig. \ref{fig1dBP} (c).  From Fig. \ref{fig1dBP}, one can see that present
experimental measurements give quite strong bounds on the differential branching ratios of $D^+\to \eta'\mu^+\nu_\mu,~D^+_s\to \pi^0\mu^+\nu_\mu$ and $D^+_s\to \pi^0\tau^+\nu_\tau$ decays  in the $C_{1}$, $C_{3}$ and $C_{4}$ cases  as well as $D^+_s\to K^0\mu^+\nu_\mu$ decays  in the $C_{1}$ and $C_{3}$ cases, and all predictions of the four  differential branching ratios in the $C_{2}$ case have large error due to the form factor choice.  Comparing with $d\mathcal{B}(D^+_s\to \pi^0\mu^+\nu_\mu)/dq^2$ in Fig. \ref{fig1dBP} (c), as shown in Fig. \ref{fig1dBP} (d), $d\mathcal{B}(D^+_s\to \pi^0\tau^+\nu_\tau)/dq^2$  is suppressed about the order of $\mathcal{O}(10^{-4})$  by $m_\tau$.

\begin{figure}[t]
\begin{center}
\includegraphics[scale=1.0]{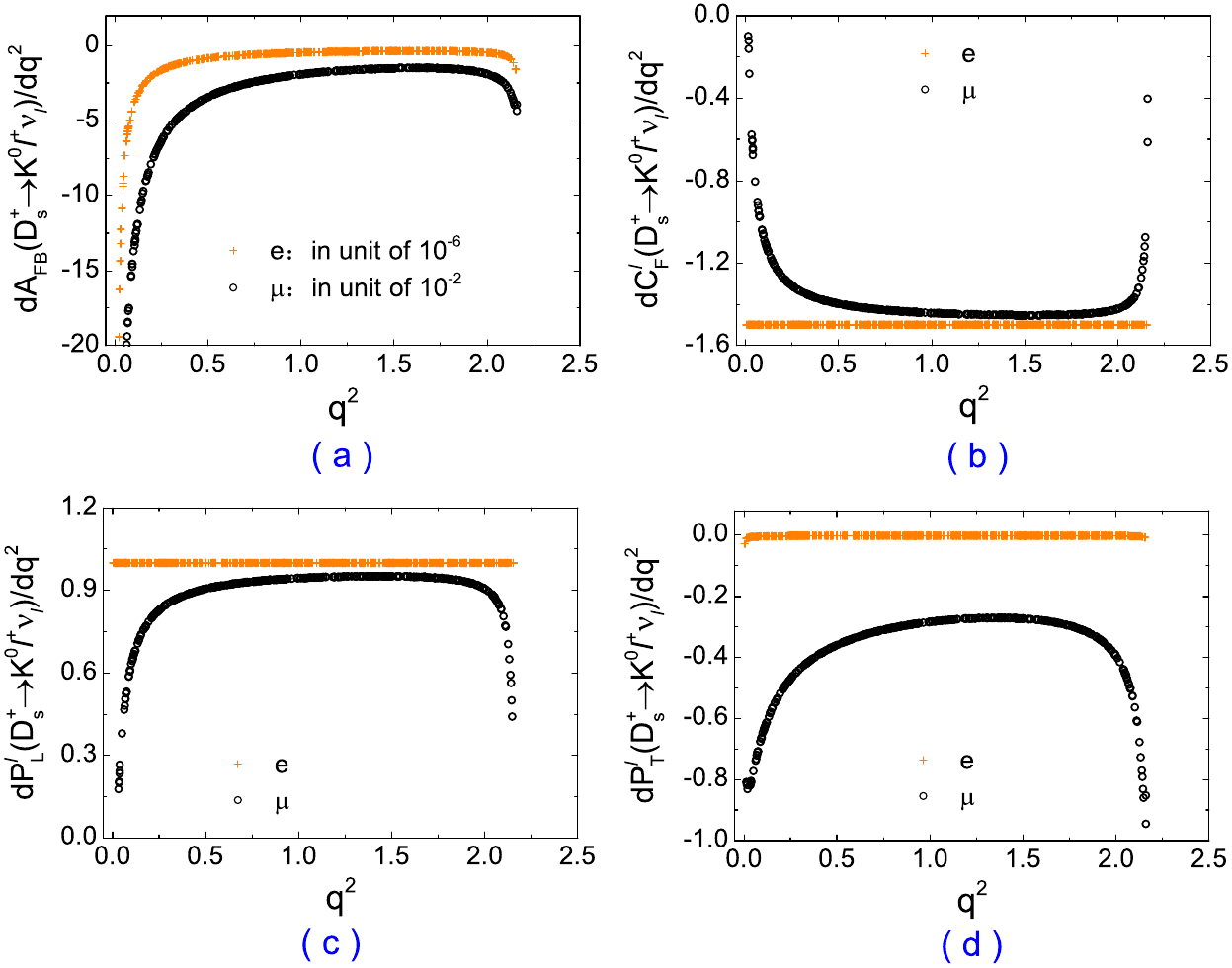}
\end{center}
\caption{ The differential forward-backward asymmetries,  differential lepton-side convexity parameters,  differential longitudinal lepton polarizations and   differential transverse lepton polarizations  for the  $D^+_s\to K^0\ell^+\nu_\ell$ decays  in the $C_3$ case. }\label{fig1dAFBP}
\end{figure}

 The forward-backward asymmetries $A^\ell_{FB}$, the lepton-side convexity parameters $ C^\ell_F$, the longitudinal polarizations of   the final charged leptons $P^\ell_L$   and the  transverse polarizations of   the final charged leptons $P^\ell_T$ with two ways of integration for the  $D\to P\ell^+\nu_\ell$ decays  could also be obtained. These predictions are very accurate, and they are similar to each other in the four $C_{1,2,3,4}$ cases. So we only give the predictions within the $C_3$ case in  Tab. \ref{Tab:AFBD2Plv} for examples.  From Tab. \ref{Tab:AFBD2Plv}, one can see that  the predictions are  obviously different  between two ways of $q^2$ integration, and the slight difference in the same way of $q^2$ integration is due to the different decay phase spaces.
 For displaying the differences between the  $D \to Pe^+\nu_e$ and $D\to P\mu^+\nu_\mu$ decays, we take $D^+_s\to K^0e^+\nu_e$ and $D^+_s\to K^0\mu^+\nu_\mu$ as examples.
 The differential forward-backward asymmetries, the  differential lepton-side convexity parameters, the  differential longitudinal lepton polarizations and the  differential transverse lepton polarizations   of $D^+_s\to K^0e^+\nu_e$ and $D^+_s\to K^0\mu^+\nu_\mu$ decays  within the $C_3$ case are displayed in Fig. \ref{fig1dAFBP}.  And one can see that  differential observables between $\ell=e$ and $\ell=\mu$ are obviously different, specially in the low and high $q^2$ ranges.

\begin{table}[t]
\renewcommand\arraystretch{1.26}
\tabcolsep 0.00in
\centering
\caption{Quantities $\langle X \rangle$ and $\overline{X}$ of the $D\to P\ell^+\nu$ in $C_3$ case. }\vspace{0.1cm}
{\footnotesize
\begin{tabular}{lcccccccc}  \hline\hline
Decay modes  & $\langle A^\ell_{FB}\rangle$& $ ^{\overline{A^e_{FB}}(\times10^{-6})}_{\overline{A^{\mu,\tau}_{FB}}(\times10^{-2})}$ & $\langle C^\ell_F\rangle$&$\overline{C^\ell_F}$&$\langle P^\ell_L\rangle$&$\overline{P^\ell_L}$&$\langle P^\ell_T\rangle$& $^{\overline{P^e_T}(\times10^{-3})}_{\overline{P^{\mu,\tau}_T}}$ \\\hline
$D^+\to \overline{K}^0e^+\nu_e    $&~~~~$-0.087$~~~~&$-3.254\pm0.001 $&$-1.239$&$-1.500 $&$0.768$&$1.000 $&$-0.273$& $ -2.442\pm0.001 $ \\
$D^+\to \pi^0e^+\nu_e             $&$-0.083$&$-2.054\pm0.000 $&$-1.252$&$-1.500 $&$0.780$&$ 1.000$&$-0.260$& $ -1.730\pm0.000 $ \\
$D^+\to \eta e^+\nu_e             $&$-0.087$&$-3.476\pm0.001 $&$-1.239$&$-1.500 $&$0.768$&$1.000 $&$-0.273$& $-2.490\pm0.000  $ \\
$D^+\to \eta'e^+\nu_e             $&$-0.093$&$-7.075\pm0.003 $&$-1.222$&$-1.500 $&$0.753$&$1.000 $&$-0.290$& $-3.890\pm0.001  $ \\
$D^0\to K^-e^+\nu_e               $&$-0.087$&$-3.259\pm0.001 $&$-1.239$&$-1.500 $&$0.768$&$1.000 $&$-0.273$& $-2.446\pm0.001  $ \\
$D^0\to \pi^-e^+\nu_e             $&$-0.083$&$-2.077\pm0.000 $&$-1.252$&$-1.500 $&$0.779$&$1.000 $&$-0.260$& $-1.751\pm0.000  $ \\
$D^+_s\to \eta e^+\nu_e           $&$-0.086$&$-3.033\pm0.001 $&$-1.242$&$-1.500 $&$0.770$&$1.000 $&$-0.270$& $-2.300\pm0.001  $ \\
$D^+_s\to \eta'e^+\nu_e           $&$-0.091$&$-5.829\pm0.003 $&$-1.226$&$-1.500 $&$0.757$&$1.000 $&$-0.286$& $-3.484\pm0.001  $ \\
$D^+_s\to K^0e^+\nu_e             $&$-0.085$&$-2.814\pm0.001 $&$-1.245$&$-1.500 $&$0.773$&$1.000 $&$-0.267$& $-2.118\pm0.000  $ \\
$D^+_s\to \pi^0e^+\nu_e           $&$-0.082$&$-1.850\pm0.001 $&$-1.254$&$-1.500 $&$0.781$&$1.000 $&$-0.258$& $-1.634\pm0.001  $ \\\hline
$D^+\to \overline{K}^0\mu^+\nu_\mu$&$-0.226$&$-4.278\pm0.001 $&$-0.822$&$-1.352 $&$0.394$&$0.851 $&$-0.655$& $ -0.414 $ \\
$D^+\to \pi^0\mu^+\nu_\mu         $&$-0.201$&$-2.810\pm0.000 $&$-0.897$&$ -1.405$&$0.462$&$0.907 $&$-0.602$& $-0.310  $ \\
$D^+\to \eta \mu^+\nu_\mu         $&$-0.227$&$-4.490\pm0.001 $&$-0.819$&$-1.347 $&$0.391$&$0.846 $&$-0.657$& $-0.419  $ \\
$D^+\to \eta'\mu^+\nu_\mu         $&$-0.263$&$-8.097\pm0.003 $&$-0.708$&$-1.213 $&$0.287$&$0.703 $&$-0.725$& $ -0.581 $ \\
$D^0\to K^-\mu^+\nu_\mu           $&$-0.226$&$-4.285\pm0.001 $&$-0.822$&$-1.352 $&$0.393$&$0.850 $&$-0.656$& $ -0.414 $ \\
$D^0\to \pi^-\mu^+\nu_\mu         $&$-0.201$&$-2.844\pm0.001 $&$-0.895$&$-1.407 $&$0.461$&$0.910 $&$-0.603$& $-0.313  $ \\
$D^+_s\to \eta \mu^+\nu_\mu       $&$-0.221$&$-4.001\pm0.001 $&$-0.836$&$-1.364 $&$0.406$&$0.864 $&$-0.646$& $-0.394  $ \\
$D^+_s\to \eta'\mu^+\nu_\mu       $&$-0.254$&$-6.952\pm0.003 $&$-0.736$&$-1.254 $&$0.314$&$0.747 $&$-0.709$& $ -0.540 $ \\
$D^+_s\to K^0\mu^+\nu_\mu         $&$-0.215$&$-3.701\pm0.001 $&$-0.856$&$-1.377 $&$0.425$&$0.879 $&$-0.632$& $-0.367 $ \\
$D^+_s\to \pi^0\mu^+\nu_\mu       $&$-0.197$&$-2.571\pm0.001 $&$-0.907$&$-1.417 $&$0.472$&$0.920 $&$-0.594$& $ -0.295 $ \\\hline
$D^+_s\to \pi^0\tau^+\nu_\tau          $&$-0.281$&$-27.429\pm0.105$&$-0.211\pm0.003$&$-0.212\pm0.003 $&$-0.868\pm0.001$&$-0.873\pm0.001 $&$-0.447\pm0.002$& $-0.437\pm0.002  $ \\
\hline
\end{tabular}\label{Tab:AFBD2Plv}}
\end{table}
%

\subsection{$D\to V\ell^+\nu_\ell$ decays}

The hadronic helicity amplitudes  for the $D\rightarrow V\ell^+\nu_\ell$  decays are   given  in  Tab. \ref{Tab:HD2VlvAmp}.
There are four  nonperturbative  parameters $B_{1,2,3,4}$ in the  $D\rightarrow V\ell^+\nu_\ell$ decay modes.  If  neglecting the SU(3) flavor breaking $c^V_1$ and $c^V_2$ terms, $B_1=B_2=B_3=B_4=c^V_0$, and then all hadronic helicity amplitudes of $D\rightarrow V\ell^+\nu_\ell$  are related by only one parameter $c^V_0$.
\begin{table}[t]
\renewcommand\arraystretch{1.26}
\tabcolsep 0.15in
\centering
\caption{The hadronic helicity  amplitudes for  $D\to V\ell^+\nu$ decays including  both the SU(3) flavor symmetry and the SU(3) flavor breaking contributions.
$B_1=c^V_0+c^V_1-2c^V_2$, $B_2=c^V_0-2c^V_1-2c^V_2$, $B_3=c^V_0+c^V_1+c^V_2$, $B_4=c^V_0-2c^V_1+c^V_2$. If neglecting the SU(3) flavor breaking $c^V_1$ and $c^V_2$ terms, $B_1=B_2=B_3=B_4=c^V_0$.  }\vspace{0.1cm}
{\footnotesize
\begin{tabular}{lccccc}  \hline\hline
Hadronic helicity amplitudes  & SU(3) IRA amplitudes\\\hline
$H(D^0\to K^{*-}\ell^+\nu_\ell)$&$B_1V^*_{cs}$\\
$H(D^+\to \overline{K}^{*0}\ell^+\nu_\ell)$&$B_1V^*_{cs}$\\
%
%
%
$H(D^+_s\to \phi\ell^+\nu_\ell)$&$\big(-cos\theta_V\sqrt{2/3}-sin\theta_V/\sqrt{3}\big)B_2V^*_{cs}$ \\
$H(D^+_s\to \omega\ell^+\nu_\ell)$&$\big(-sin\theta_V\sqrt{2/3}+cos\theta_V/\sqrt{3}\big)B_2V^*_{cs}$ \\\hline
$H(D^0\to \rho^-\ell^+\nu_\ell)$&$B_3V^*_{cd}$ \\
$H(D^+\to \rho^0\ell^+\nu_\ell)$&$-\frac{1}{\sqrt{2}}B_3V^*_{cd}$ \\
%
%
%
$H(D^+\to \phi\ell^+\nu_\ell)$&$\big(cos\theta_V/\sqrt{6}-sin\theta_V/\sqrt{3}\big)B_3V^*_{cd}$ \\
$H(D^+\to \omega\ell^+\nu_\ell)$&$\big(sin\theta_V/\sqrt{6}+cos\theta_V/\sqrt{3}\big)B_3V^*_{cd}$  \\
$H(D^+_s\to K^{*0}\ell^+\nu_\ell)$&$B_4V^*_{cd}$ \\\hline
\end{tabular}\label{Tab:HD2VlvAmp}}
\end{table}

Among the $D\rightarrow V\ell^+\nu_\ell$ decay modes,   13 branching ratios have been measured, and  2 branching ratios have been upper limited by the experiments.
 The experimental data with $2\sigma$ errors are listed in the second column of Tab. \ref{Tab:BrD2Vlv}.
Now  we  use the listed experimental data to  constrain the  parameters $B_i$  and then predict other not yet measured and  not yet well measured  branching ratios.
 The numerical results of $\mathcal{B}(D\to V\ell^+\nu_\ell)$ in the $C_{1}$, $C_{2}$, $C_{3}$ and $C_{4}$ cases  are given in the third, forth, fifth and sixth columns of Tab. \ref{Tab:BrD2Vlv}, respectively.
\begin{table}[ht]
\renewcommand\arraystretch{1.22}
\tabcolsep 0.18in
\centering
\caption{Branching ratios of the $D\to V\ell^+\nu$ within $2\sigma$ errors.  $^\dag$The experimental data of  $\mathcal{B}(D^+\to \omega e^+\nu_e)$  and $\mathcal{B}(D^0\to \rho^- \mu^+\nu_\mu)$  from PDG \cite{PDG2022} are not used in the $C_{1,2,3}$  cases.  }\vspace{0.1cm}
{\footnotesize
\begin{tabular}{lccccccc}  \hline\hline
Branching ratios                                                 &    Exp. data               &    Ones in  $C_1$      &     Ones in $C_2$    &   Ones in $C_3$      &  Ones in $C_4$    \\\hline
$\mathcal{B}(D^+\to \overline{K}^{*0}e^+\nu_e)(\times10^{-2})$   &   $5.40\pm0.20$            &   $5.44\pm0.15$        &   $5.42\pm0.18$      &   $5.36\pm0.08$      &   $5.44\pm0.16$    \\
$\mathcal{B}(D^+\to \rho^0e^+\nu_e)(\times10^{-3})$              &   $2.18^{+0.34}_{-0.50}$   &   $2.31\pm0.07$        &   $2.39\pm0.13$      &   $2.33\pm0.05$      &   $1.83\pm0.15$     \\
$\mathcal{B}(D^+\to \omega e^+\nu_e)(\times10^{-3})$             &   $1.69\pm0.22$            &   $2.24\pm0.07^\dag$   &   $2.33\pm0.12^\dag$ &   $2.26\pm0.04^\dag$ &   $1.77\pm0.14$    \\
$\mathcal{B}(D^+\to \phi e^+\nu_e)(\times10^{-7})$               &   $<130$                   &   $3.13\pm0.12$        &   $3.11\pm0.19$      &   $3.07\pm0.07$      &   $2.38\pm0.23$     \\
$\mathcal{B}(D^0\to K^{*-}e^+\nu_e)(\times10^{-2})$              &   $2.15\pm0.32$            &   $2.12\pm0.09$        &   $2.13\pm0.10$      &   $2.08\pm0.06$      &   $2.13\pm0.10$     \\
$\mathcal{B}(D^0\to \rho^-e^+\nu_e)(\times10^{-3})$              &   $1.50\pm0.24$            &   $1.79\pm0.08$        &   $1.86\pm0.11$      &   $1.80\pm0.06$      &   $1.41\pm0.13$     \\
$\mathcal{B}(D^+_s\to \phi e^+\nu_e)(\times10^{-2})$             &   $2.39\pm0.32$            &   $2.46\pm0.12$        &   $2.43\pm0.14$      &   $2.40\pm0.10$      &   $2.39\pm0.32$     \\
$\mathcal{B}(D^+_s\to \omega e^+\nu_e)(\times10^{-5})$           &   $<200$                   &   $2.45\pm0.13$        &   $2.56\pm0.20$      &   $2.47\pm0.10$      &   $2.49\pm0.38$     \\
$\mathcal{B}(D^+_s\to K^{*0}e^+\nu_e)(\times10^{-3})$            &   $2.15\pm0.56$            &   $2.17\pm0.10$        &   $2.25\pm0.13$      &   $2.17\pm0.08$      &   $2.15\pm0.56$     \\\hline
$\mathcal{B}(D^+\to \overline{K}^{*0}\mu^+\nu_\mu)(\times10^{-2})$   &  $5.27\pm0.30$         &   $5.12\pm0.15$        &   $5.13\pm0.16$      &   $5.05\pm0.08$      &   $5.12\pm0.15$     \\
$\mathcal{B}(D^+\to \rho^0\mu^+\nu_\mu)(\times10^{-3})$              &   $2.4\pm0.8$          &   $2.19\pm0.07$        &   $2.29\pm0.13$      &   $2.22\pm0.04$      &   $1.74\pm0.14$     \\
$\mathcal{B}(D^+\to \omega \mu^+\nu_\mu)(\times10^{-3})$             &  $1.77\pm0.42$         &   $2.13\pm0.06$        &   $2.23\pm0.12$      &   $2.15\pm0.04$      &   $1.68\pm0.13$
\\
$\mathcal{B}(D^+\to \phi \mu^+\nu_\mu)(\times10^{-7})$               &   $\cdots$             &   $2.89\pm0.11$        &   $2.89\pm0.17$      &   $2.84\pm0.07$      &    $2.20\pm0.21$    \\
$\mathcal{B}(D^0\to K^{*-}\mu^+\nu_\mu)(\times10^{-2})$              &   $1.89\pm0.48$        &   $1.99\pm0.09$        &   $2.01\pm0.09$      &   $1.96\pm0.06$      &   $2.01\pm0.10$     \\
$\mathcal{B}(D^0\to \rho^-\mu^+\nu_\mu)(\times10^{-3})$              &   $1.35\pm0.26$        &   $1.70\pm0.07^\dag$   &   $1.78\pm0.11^\dag$ &   $1.72\pm0.06^\dag$ &   $1.34\pm0.13$     \\
$\mathcal{B}(D^+_s\to \phi \mu^+\nu_\mu)(\times10^{-2})$             &   $1.9\pm1.0$          &   $2.30\pm0.12$        &   $2.29\pm0.12$      &   $2.25\pm0.09$      &   $2.24\pm0.30$     \\
$\mathcal{B}(D^+_s\to \omega \mu^+\nu_\mu)(\times10^{-5})$           &   $\cdots$             &   $2.34\pm0.12$        &   $2.47\pm0.19$      &   $2.37\pm0.09$      &   $2.38\pm0.36$     \\
$\mathcal{B}(D^+_s\to K^{*0}\mu^+\nu_\mu)(\times10^{-3})$            &   $\cdots$             &   $2.06\pm0.10$        &   $2.15\pm0.13$      &   $2.07\pm0.08$      &     $2.05\pm0.53$    \\\hline
$R^{\mu/e}(D^+\to \overline{K}^{*0}\ell^+\nu_\ell)$                  &      $$               &    $0.939\pm0.001$      &   $0.944\pm0.004$    &   $0.941\pm0.001$   &   $0.941\pm0.001$  \\
$R^{\mu/e}(D^+\to \rho^0\ell^+\nu_\ell)$                             &      $$               &    $0.950\pm0.001$      &   $0.956\pm0.005$    &   $0.952\pm0.001$   &   $0.952\pm0.001$  \\
$R^{\mu/e}(D^+\to \omega \ell^+\nu_\ell)$                            &      $$               &    $0.950\pm0.001$      &   $0.956\pm0.005$    &   $0.952\pm0.001$   &   $0.952\pm0.001$  \\
$R^{\mu/e}(D^+\to \phi \ell^+\nu_\ell)$                              &      $$               &    $0.923\pm0.001$      &   $0.928\pm0.005$    &   $0.925\pm0.001$   &   $0.925\pm0.001$  \\
$R^{\mu/e}(D^0\to K^{*-}\ell^+\nu_\ell)$                             &      $$               &    $0.939\pm0.001$      &   $0.944\pm0.004$    &   $0.941\pm0.001$   &   $0.941\pm0.001$  \\
$R^{\mu/e}(D^0\to \rho^-\ell^+\nu_\ell)$                             &      $$               &    $0.950\pm0.001$      &   $0.956\pm0.005$    &   $0.952\pm0.001$   &   $0.952\pm0.001$  \\
$R^{\mu/e}(D^+_s\to \phi \ell^+\nu_\ell)$                            &      $$               &    $0.937\pm0.001$      &   $0.942\pm0.004$    &   $0.939\pm0.001$   &   $0.939\pm0.001$  \\
$R^{\mu/e}(D^+_s\to \omega \ell^+\nu_\ell)$                          &      $$               &    $0.957\pm0.001$      &   $0.963\pm0.004$    &   $0.959\pm0.001$   &   $0.959\pm0.001$  \\
$R^{\mu/e}(D^+_s\to K^{*0}\ell^+\nu_\ell)$                           &      $$               &    $0.949\pm0.001$      &   $0.955\pm0.005$    &   $0.951\pm0.001$   &   $0.951\pm0.001$  \\\hline
\end{tabular}\label{Tab:BrD2Vlv}}
\end{table}

The results in the $C_{1}$, $C_{2}$ and  $C_{3}$ cases    are very similar.   Since the SU(3) flavor symmetry predictions of  $\mathcal{B}(D^+\to \omega e^+\nu_e)$  and $\mathcal{B}(D^0\to \rho^- \mu^+\nu_\mu)$ are slightly larger than their experimental data within $2\sigma$ errors in the three cases,  we do not use them to constrain the nonperturbative  parameter $c^V_0$.  One can see that the prediction of $\mathcal{B}(D^0\to \rho^- \mu^+\nu_\mu)$ is agree with its experimental data within $3\sigma$ errors, nevertheless, the prediction of $\mathcal{B}(D^+\to \omega e^+\nu_e)$ still slightly larger than   experimental data within $3\sigma$ errors.   $\mathcal{B}(D^+_s\to K^{*0} \mu^+\nu_\mu)$ and $\mathcal{B}(D^+_s\to \omega e^+\nu_e,\omega \mu^+\nu_\mu)$ are predicted on the order of $\mathcal{O}(10^{-3})$ and $\mathcal{O}(10^{-5})$, respectively. And they could be measured  in BESIII, LHCb and BelleII experiments.
In the $C_{4}$ case, as given in the sixth column of Tab. \ref{Tab:BrD2Vlv},   after considering both the hadronic momentum-transfer $q^2$  dependence of the form factors and the SU(3) flavor breaking contributions,   all SU(3) flavor symmetry predictions are consistent with their experimental data within $2\sigma$ errors.
Among  relevant not yet measured  decays, $\mathcal{B}(D^+_s\to K^{*0}\mu^+\nu_\mu)$ is calculated in the SM using light-cone sum rules \cite{Leng:2020fei} and in the relativistic quark model \cite{Faustov:2019mqr},
$\mathcal{B}(D^+_s\to K^{*0}\mu^+\nu_\mu)= (2.23\pm0.32)\times10^{-3} ~\mbox{\cite{Leng:2020fei}}~\mbox{and}~2.0\times10^{-3} ~\mbox{\cite{Faustov:2019mqr}}$,
 and our predictions of $\mathcal{B}(D^+_s\to K^{*0}\mu^+\nu_\mu)$  in  the $C_{1}$, $C_{2}$, $C_{3}$ and $C_{4}$ cases are coincident with previous ones in Refs. \cite{Leng:2020fei,Faustov:2019mqr}.
In addition, the lepton flavor universality parameters $R^{\mu/e}(D\to V\ell^+\nu_\ell)$  are also given in Tab. \ref{Tab:BrD2Vlv}. Since many terms are canceled in the ratios, these predictions of the lepton flavor universality parameters are quite accurate, and our predictions in all four cases are similar to each other.

\begin{figure}[t]
\begin{center}
\includegraphics[scale=0.95]{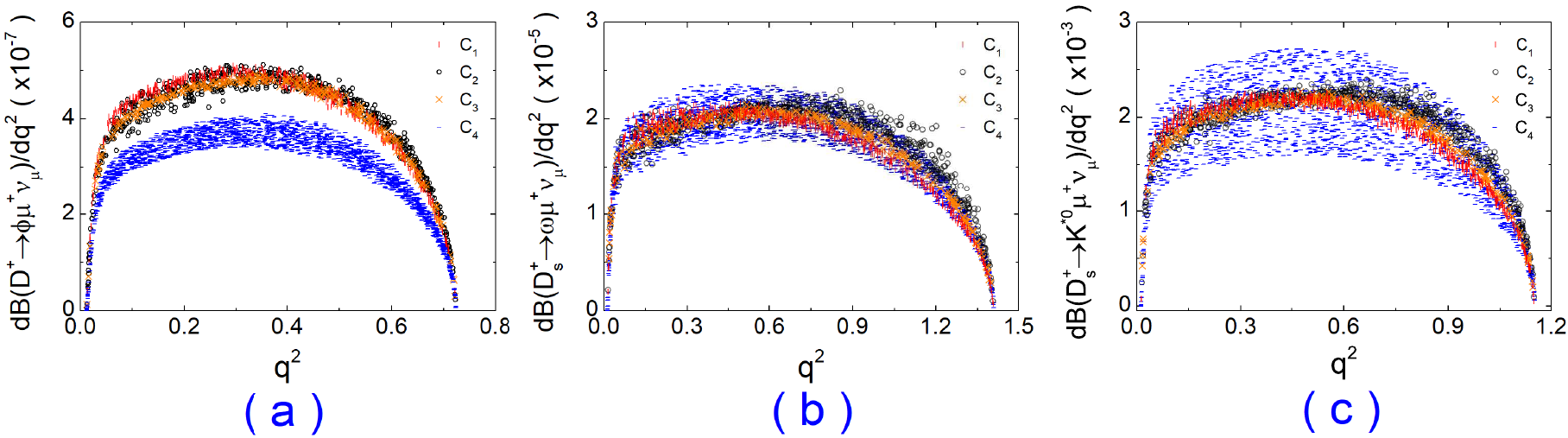}
\end{center}
\caption{The $q^2$ dependence of the differential branching ratios for some not yet measured  $D\to V\mu^+\nu_\mu$ decays  with present experimental bounds. }\label{fig3dBV}
%
\begin{center}
\includegraphics[scale=1.1]{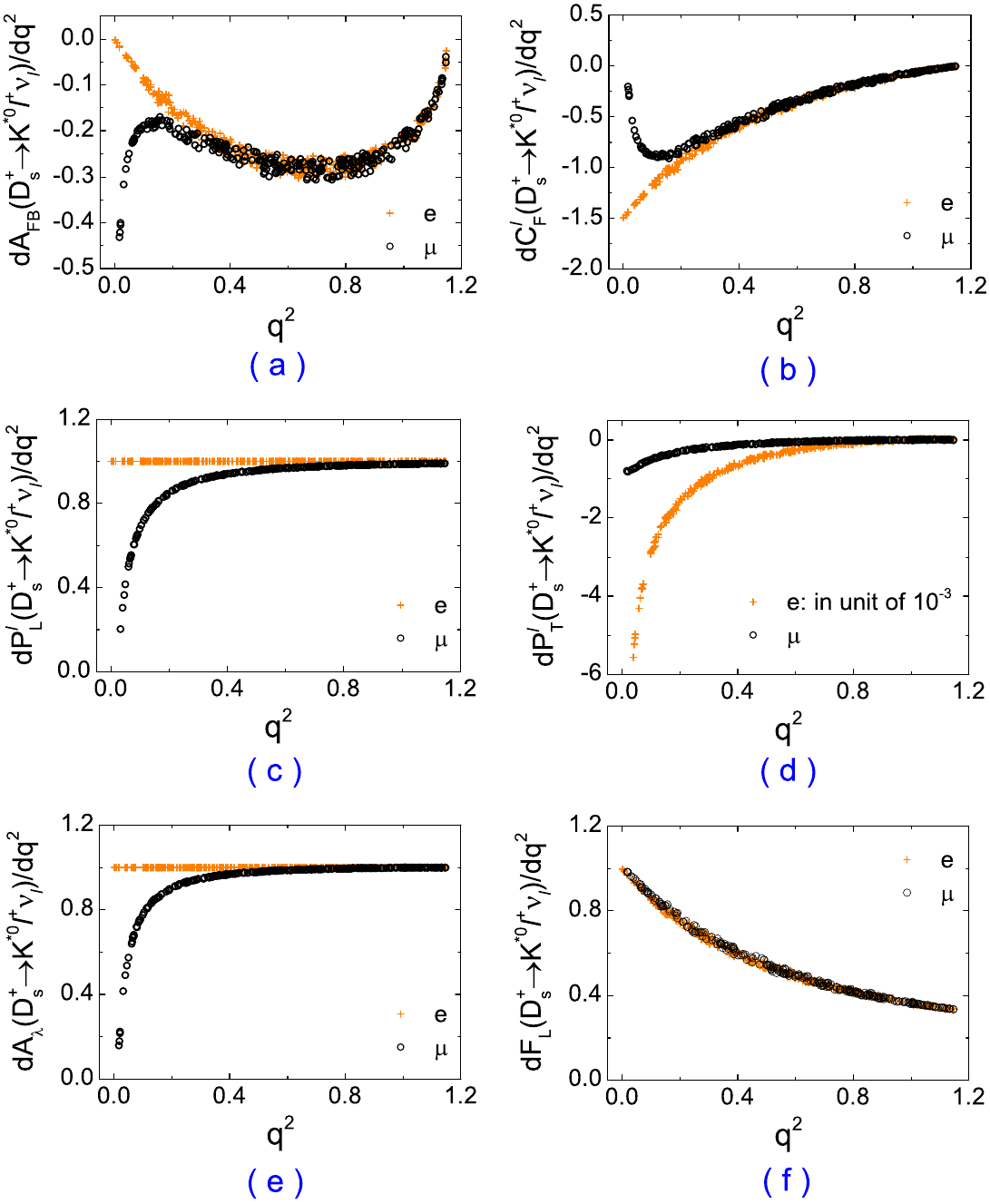}
\end{center}
\caption{The differential forward-backward asymmetries,  differential lepton-side convexity parameters,  differential longitudinal lepton polarizations and   differential transverse lepton polarizations  for the $D^+_s\to K^0\ell^+\nu_\ell$ decays  in the $C_3$ case. }\label{fig4dAFBV}
\end{figure}

For the $q^2$ dependence of the differential branching ratios of the $D\to V\ell^+\nu_\ell$ decays with present experimental bounds, we only show  the not yet measured  processes $D^+\to \phi\mu^+\nu_\mu, ~D^+_s\to \omega\mu^+\nu_\mu$ and $D^+_s\to K^{*0}\mu^+\nu_\mu$ in Fig. \ref{fig3dBV}.   The differential branching ratios of $D^+\to \phi e^+\nu_e~(D^+_s\to \omega e^+\nu_e)$ is similar to $D^+\to \phi\mu^+\nu_\mu~(D^+_s\to \omega\mu^+\nu_\mu)$, so we do not shown them in  Fig. \ref{fig3dBV}.
 From Fig. \ref{fig3dBV}, one can see that present
experiment data give quite strong bounds on all differential branching ratios of $D^+\to \phi\mu^+\nu_\mu, ~D^+_s\to \omega\mu^+\nu_\mu$ and $D^+_s\to K^{*0}\mu^+\nu_\mu$ decays  in the $C_{1}$, $C_{2}$ and $C_{3}$ cases. The prediction of $d\mathcal{B}(D^+\to \phi\mu^+\nu_\mu)/dq^2$ in the $C_4$ case could be distinguished from ones in the $C_{1,2,3}$ cases  within the middle range of $q^2$. And  the error of  $d\mathcal{B}(D^+_s\to K^{*0}\mu^+\nu_\mu)/dq^2$ in the $C_4$ case is obviously  larger than ones in $C_{1,2,3}$ cases.

The forward-backward asymmetries $A^\ell_{FB}$, the lepton-side convexity parameters $ C^\ell_F$, the longitudinal polarizations $P^\ell_L$, the transverse polarizations $P^\ell_T$,  the lepton spin asymmetries $A_{\lambda}$ and  the longitudinal polarization fractions of the final vector mesons $F_L$  with two ways of integration  have also been predicted in the four cases. Since many theoretical uncertainties are canceled in the ratios, these predictions are very accurate. These predictions are similar to each other in the four cases, and we only list the results in the $C_3$ case  in  Tabs. \ref{Tab:D2VlvXa}-\ref{Tab:D2VlvXb} for examples.  One can see that  the predictions are  obviously different  between two ways of $q^2$ integration, and they are also quite different between $D\to Ve^+\nu_e$  and $D\to V\mu^+\nu_\mu$ decays.
\begin{table}[b]
\renewcommand\arraystretch{1.2}
\tabcolsep 0.12in
\centering
\caption{The forward-backward asymmetries $A^\ell_{FB}$, the lepton-side convexity parameters $ C^\ell_F$, the longitudinal polarizations $P^\ell_L$  of the $D\to V\ell^+\nu$ decays in the $C_3$ case. }\vspace{0.1cm}
{\footnotesize
\begin{tabular}{lcccccccccccc}  \hline\hline
Decay modes     &    $
\langle A^\ell_{FB}\rangle$          &$\overline{ A^\ell_{FB}}$ &    $\langle C^\ell_F\rangle$    &   $\overline{C^\ell_F}$   &   $\langle P^\ell_L\rangle$   &   $\overline{P^\ell_L}$\\\hline
$D^+\to \overline{K}^{*0}e^+\nu_e$       &   $-0.125\pm0.006$   &   $-0.190\pm0.020$   &   $-1.046\pm0.019$   &   $-0.500\pm0.032$   &   $0.786\pm0.004$   &   $1.000$ \\
$D^+\to \rho^0e^+\nu_e           $       &   $-0.130\pm0.008$   &   $-0.222\pm0.024$   &   $-1.052\pm0.023$   &   $-0.496\pm0.041$   &   $0.789\pm0.004$   &   $1.000$ \\
$D^+\to \omega e^+\nu_e          $       &   $-0.130\pm0.008$   &   $-0.220\pm0.024$   &   $-1.052\pm0.023$   &   $-0.497\pm0.041$   &   $0.789\pm0.004$   &   $1.000$ \\
$D^+\to \phi e^+\nu_e            $       &   $-0.121\pm0.005$   &   $-0.164\pm0.017$   &   $-1.037\pm0.015$   &   $-0.500\pm0.025$   &   $0.784\pm0.003$   &   $1.000$ \\
$D^0\to K^{*-}e^+\nu_e           $       &   $-0.125\pm0.006$   &   $-0.191\pm0.020$   &   $-1.046\pm0.019$   &   $-0.500\pm0.032$   &   $0.786\pm0.004$   &   $1.000$ \\
$D^0\to \rho^-e^+\nu_e           $       &   $-0.130\pm0.008$   &   $-0.221\pm0.024$   &   $-1.052\pm0.023$   &   $-0.497\pm0.041$   &   $0.789\pm0.004$   &   $1.000$ \\
$D^+_s\to \phi e^+\nu_e          $       &   $-0.122\pm0.006$   &   $-0.176\pm0.018$   &   $-1.043\pm0.016$   &   $-0.500\pm0.028$   &   $0.786\pm0.003$   &   $1.000$ \\
$D^+_s\to \omega e^+\nu_e        $       &   $-0.130\pm0.008$   &   $-0.229\pm0.025$   &   $-1.057\pm0.025$   &   $-0.496\pm0.044$   &   $0.790\pm0.004$   &   $1.000$ \\
$D^+_s\to K^{*0}e^+\nu_e         $       &   $-0.128\pm0.007$   &   $-0.207\pm0.022$   &   $-1.049\pm0.021$   &   $-0.495\pm0.036$   &   $0.789\pm0.004$   &   $1.000$ \\\hline
$D^+\to \overline{K}^{*0}\mu^+\nu_\mu$   &   $-0.284\pm0.009$   &   $-0.226\pm0.019$   &   $-0.466\pm0.021$   &   $-0.395\pm0.028$   &   $0.514\pm0.017$   &   $0.886\pm0.002$\\
$D^+\to \rho^0\mu^+\nu_\mu           $   &   $-0.292\pm0.011$   &   $-0.252\pm0.023$   &   $-0.491\pm0.027$   &   $-0.405\pm0.037$   &   $0.524\pm0.020$   &   $0.903\pm0.002$\\
$D^+\to \omega \mu^+\nu_\mu          $   &   $-0.292\pm0.011$   &   $-0.251\pm0.022$   &   $-0.490\pm0.027$   &   $-0.405\pm0.037$   &   $0.524\pm0.020$   &   $0.902\pm0.002$\\
$D^+\to \phi \mu^+\nu_\mu            $   &   $-0.277\pm0.008$   &   $-0.206\pm0.016$   &   $-0.433\pm0.016$   &   $-0.376\pm0.021$   &   $0.503\pm0.014$   &   $0.864\pm0.002$\\
$D^0\to K^{*-}\mu^+\nu_\mu           $   &   $-0.284\pm0.009$   &   $-0.226\pm0.019$   &   $-0.466\pm0.021$   &   $-0.395\pm0.029$   &   $0.514\pm0.017$   &   $0.886\pm0.002$\\
$D^0\to \rho^-\mu^+\nu_\mu           $   &   $-0.292\pm0.011$   &   $-0.252\pm0.023$   &   $-0.490\pm0.027$   &   $-0.405\pm0.037$   &   $0.524\pm0.020$   &   $0.902\pm0.002$\\
$D^+_s\to \phi \mu^+\nu_\mu          $   &   $-0.277\pm0.008$   &   $-0.213\pm0.017$   &   $-0.459\pm0.018$   &   $-0.391\pm0.024$   &   $0.514\pm0.015$   &   $0.882\pm0.002$\\
$D^+_s\to \omega \mu^+\nu_\mu        $   &   $-0.291\pm0.012$   &   $-0.257\pm0.024$   &   $-0.509\pm0.029$   &   $-0.414\pm0.041$   &   $0.531\pm0.021$   &   $0.913\pm0.002$\\
$D^+_s\to K^{*0}\mu^+\nu_\mu         $   &   $-0.286\pm0.010$   &   $-0.239\pm0.021$   &   $-0.485\pm0.024$   &   $-0.402\pm0.033$   &   $0.525\pm0.018$   &   $0.900\pm0.002$\\\hline
\end{tabular}\label{Tab:D2VlvXa}}
%
\renewcommand\arraystretch{1.2}
\tabcolsep 0.12in
\centering
\caption{The transverse polarizations $P^\ell_T$, the lepton spin asymmetries $A_{\lambda}$ and  the longitudinal polarization fractions of the final vector mesons $F_L$  of the $D\to V\ell^+\nu$ decays in the $C_3$ case. }\vspace{0.1cm}
{\footnotesize
\begin{tabular}{lcccccccccccc}  \hline\hline
Decay modes     &    $\langle P^\ell_T\rangle$   &   $  ^{\overline{P^e_T}(\times10^{-3})}_{\overline{P^\mu_T}}$   &   $\langle A_{\lambda}\rangle$    &   $\overline{A_{\lambda}}$   &   $\langle F_L\rangle$   &   $\overline{F_L}$ \\\hline
$D^+\to \overline{K}^{*0}e^+\nu_e$       &   $-0.251\pm0.004 $   &   $-1.205\pm0.066 $   &        $ 1.000$       &   $1.000 $           &   $ 0.905\pm0.010$    &   $0.556\pm0.014 $\\
$D^+\to \rho^0e^+\nu_e           $       &   $-0.249\pm0.005 $   &   $-1.040\pm0.072 $   &        $ 1.000$       &   $1.000  $          &   $ 0.907\pm0.012$    &   $0.554\pm0.018 $\\
$D^+\to \omega e^+\nu_e          $       &   $-0.249\pm0.005 $   &   $-1.049\pm0.073 $   &        $1.000 $       &   $1.000  $          &   $ 0.907\pm0.012$    &   $0.554\pm0.018 $\\
$D^+\to \phi e^+\nu_e            $       &   $-0.254\pm0.003 $   &   $-1.417\pm0.061 $   &        $1.000 $       &   $1.000  $          &   $0.902\pm0.008 $    &   $0.556\pm0.011 $\\
$D^0\to K^{*-}e^+\nu_e           $       &   $-0.251\pm0.004 $   &   $-1.206\pm0.067 $   &        $1.000 $       &   $1.000  $          &   $0.905\pm0.010 $    &   $0.556\pm0.014 $\\
$D^0\to \rho^-e^+\nu_e           $       &   $-0.249\pm0.005 $   &   $-1.045\pm0.073 $   &        $1.000 $       &   $1.000 $           &   $0.907\pm0.012 $    &   $0.554\pm0.018 $\\
$D^+_s\to \phi e^+\nu_e          $       &   $-0.251\pm0.004 $   &   $-1.255\pm0.060 $   &        $1.000 $       &   $1.000  $          &   $0.904\pm0.009 $    &   $0.555\pm0.012 $\\
$D^+_s\to \omega e^+\nu_e        $       &   $-0.247\pm0.005 $   &   $-0.953\pm0.071 $   &        $1.000 $       &   $1.000  $          &   $0.908\pm0.013 $    &   $0.554\pm0.020 $\\
$D^+_s\to K^{*0}e^+\nu_e         $       &   $-0.248\pm0.004 $   &   $-1.075\pm0.066 $   &        $1.000 $       &   $1.000  $          &   $0.905\pm0.011 $    &   $0.553\pm0.016$\\\hline
$D^+\to \overline{K}^{*0}\mu^+\nu_\mu$   &   $-0.454\pm0.022 $     &   $-0.156\pm0.012 $     &   $0.935\pm0.005 $   &   $0.928\pm0.002 $   &   $0.775\pm0.019 $   &   $0.557\pm0.014 $\\
$D^+\to \rho^0\mu^+\nu_\mu           $   &   $-0.452\pm0.026 $     &   $-0.139\pm0.014 $     &   $0.944\pm0.006 $   &   $0.937\pm0.002 $   &   $0.782\pm0.023 $   &   $0.555\pm0.018 $\\
$D^+\to \omega \mu^+\nu_\mu          $   &   $-0.452\pm0.026 $     &   $-0.140\pm0.014 $     &   $0.944\pm0.006 $   &   $0.937\pm0.002 $   &   $0.782\pm0.023 $   &   $0.555\pm0.018 $\\
$D^+\to \phi \mu^+\nu_\mu            $   &   $-0.455\pm0.018 $     &   $-0.175\pm0.011 $     &   $0.924\pm0.005 $   &   $0.915\pm0.002 $   &   $0.763\pm0.015 $   &   $0.557\pm0.011 $\\
$D^0\to K^{*-}\mu^+\nu_\mu           $   &   $-0.454\pm0.022 $     &   $-0.156\pm0.012 $     &   $0.935\pm0.005 $   &   $0.927\pm0.002 $   &   $0.775\pm0.019 $   &   $0.557\pm0.014 $\\
$D^0\to \rho^-\mu^+\nu_\mu           $   &   $-0.452\pm0.026 $     &   $-0.140\pm0.014 $     &   $0.944\pm0.006 $   &   $0.937\pm0.002 $   &   $0.782\pm0.023 $   &   $0.555\pm0.018 $\\
$D^+_s\to \phi \mu^+\nu_\mu          $   &   $-0.454\pm0.019 $     &   $-0.162\pm0.011 $     &   $0.934\pm0.005 $   &   $0.925\pm0.002 $   &   $0.771\pm0.016 $   &   $0.557\pm0.012 $\\
$D^+_s\to \omega \mu^+\nu_\mu        $   &   $-0.452\pm0.027 $     &   $-0.131\pm0.014 $     &   $0.950\pm0.005 $   &   $0.943\pm0.002 $   &   $ 0.788\pm0.024$   &   $0.555\pm0.019 $\\
$D^+_s\to K^{*0}\mu^+\nu_\mu         $   &   $-0.451\pm0.023 $     &   $-0.143\pm0.012 $     &   $0.943\pm0.005 $   &   $0.936\pm0.002 $   &   $0.779\pm0.021 $   &   $0.555\pm0.016 $\\\hline
\end{tabular}\label{Tab:D2VlvXb}}
\end{table}

The differential observables  of $D^+_s\to K^{*0}\ell^+\nu_\ell$ decays  in the $C_3$ case are displayed in Fig. \ref{fig4dAFBV}. One can see that, in the low  $q^2$ ranges,   the differential observables expect $dF_L(D^+_s\to K^{*0}\ell^+\nu_\ell)/dq^2$ are obviously different  between decays with $\ell=e$ and $\ell=\mu$.

\clearpage
\subsection{$D\to S\ell^+\nu_\ell$ decays}

For $D\to S\ell^+\nu_\ell$ decays, the two quark and the four quark scenarios for the scalar mesons below or near 1 $GeV$  are considered.
The hadronic helicity amplitudes  for the $D\rightarrow S\ell^+\nu_\ell$  decays are   given  in  Tab. \ref{Tab:HD2SlvAmp}, in which   the CKM matrix element $V_{cs}$ and $V_{cd}$ information are kept  for comparing conveniently.  There are four (five)  nonperturbative  parameters $E_{1,2,3,4}$ ($E'_{1,2,3,4,5}$) in the two quark (four quark) picture.
After ignoring the SU(3) flavor breaking contributions,  only one nonperturbative  parameter  $E_1=E_2=E_3=E_4=c^S_0$ or $E'_1=E'_2=E'_3=E'_4=E'_5=c'^S_0$  relates all decay amplitudes in  the two quark or the four quark picture,  respectively.

\begin{table}[b]
\renewcommand\arraystretch{1.4}
\tabcolsep 0.2in
\centering
\caption{The hadronic helicity  amplitudes for  $D\to S\ell^+\nu$ decays including  both the SU(3) flavor symmetry and the SU(3) flavor breaking contributions.
In the two quark picture of the scalar mesons, $E_1\equiv c^S_0+c^S_1-2c^S_2$, $E_2\equiv c^S_0-2c^S_1-2c^S_2$, $E_3\equiv c^S_0+c^S_1+c^S_2$, $E_4\equiv c^S_0-2c^S_1+c^S_2$.  $E_1=E_2=E_3=E_4=c^S_0$ if  neglecting the SU(3) flavor breaking $c^S_1$ and $c^S_2$ terms.
In the four quark picture of the scalar mesons, $E'_1\equiv c'^S_0+c'^S_1-2c'^S_2+c'^S_3$,   $E'_2\equiv c'^S_0-2c'^S_1-2c'^S_2+c'^S_3$, $E'_3\equiv c'^S_0+c'^S_1+c'^S_2-2c'^S_3$,  $E'_4\equiv c'^S_0+c'^S_1+c'^S_2+c'^S_3$, $E'_5\equiv c'^S_0-2c'^S_1+c'^S_2+c'^S_3$, $E'_1=E'_2=E'_3=E'_4=E'_5=c'^S_0$ if  neglecting the SU(3) flavor breaking $c'^S_1$, $c'^S_2$ and $c'^S_3$ terms. }\vspace{0.1cm}
\begin{tabular}{lcc}  \hline\hline
Hadronic helicity amplitudes  & ones for two-quark scenario & ones for four-quark scenario \\\hline
$H(D^0\to K^-_0\ell^+\nu_\ell)$&$E_1V^*_{cs}$&$E'_1V^*_{cs}$\\
$H(D^+\to \overline{K}_0^0\ell^+\nu_\ell)$&$E_1V^*_{cs}$&$E'_1V^*_{cs}$\\
$H(D^+_s\to f_0\ell^+\nu_\ell)$&$E_2V^*_{cs}$ &$\sqrt{2}E'_2V^*_{cs}$ \\
$H(D^+_s\to f_0(980)\ell^+\nu_\ell)$&$cos\theta_S~E_2V^*_{cs}$&$\sqrt{2}cos\phi_S~E'_2V^*_{cs}$  \\
$H(D^+_s\to f_0(500)\ell^+\nu_\ell)$&$-sin\theta_S~E_2V^*_{cs}$ &$-\sqrt{2}sin\phi_S~E'_2V^*_{cs}$\\
\hline
$H(D^0\to a^-_0\ell^+\nu_\ell)$&$E_3V^*_{cd}$&$E'_3V^*_{cd}$\\
$H(D^+\to a^0_0\ell^+\nu_\ell)$&$-\frac{1}{\sqrt{2}}E_3V^*_{cd}$&$-\frac{1}{\sqrt{2}}E'_3V^*_{cd}$ \\
$H(D^+\to f_0 \ell^+\nu_\ell)$&$0$&$\frac{1}{\sqrt{2}}E'_3V^*_{cd}$ \\%
$H(D^+\to \sigma \ell^+\nu_\ell)$&$\frac{1}{\sqrt{2}}E_3V^*_{cd}$&$E'_4V^*_{cd}$ \\
$H(D^+\to f_0(980) \ell^+\nu_\ell)$&$\frac{1}{\sqrt{2}}sin\theta_S~E_3V^*_{cd}$&$(\frac{1}{\sqrt{2}}E'_3cos\phi_S+E'_4sin\phi_S)V^*_{cd}$ \\
$H(D^+\to f_0(500) \ell^+\nu_\ell)$&$\frac{1}{\sqrt{2}}cos\theta_S~E_3V^*_{cd}$&$(-\frac{1}{\sqrt{2}}E'_3sin\phi_S+E'_4cos\phi_S)V^*_{cd}$ \\
$H(D^+_s\to K^0_0\ell^+\nu_\ell)$&$E_4V^*_{cd}$&$E'_5V^*_{cd}$  \\\hline
\end{tabular}\label{Tab:HD2SlvAmp}
\end{table}

Unlike many measured decay modes in the $D\rightarrow P\ell^+\nu_\ell$  and  $D\rightarrow V\ell^+\nu_\ell$ decays, among these  $D\rightarrow S\ell^+\nu_\ell$  decays, only $D^+_s\to f_0(980)e^+\nu_e$ decay has been measured, and its branching ratio with $2\sigma$ errors is  \cite{PDG2022}
\begin{eqnarray}
\mathcal{B}(D^+_s\to f_0(980)e^+\nu_e)=(2.3\pm0.8)\times10^{-3}. \label{Eq:BrDs2f0980ev}
\end{eqnarray}
In addition, the branching ratios of the $D\to P_1P_2\ell^+\nu_\ell$  decays with the light scalar resonances can be obtained by using $\mathcal{B}(D\to S\ell^+\nu_\ell)$ and $\mathcal{B}(S\to P_1P_2)$, and the detail analysis can been found in Ref. \cite{wang:2022ourwork}.  Five branching ratios and two upper limits of $\mathcal{B}(D\to S\ell^+\nu_\ell,~S\to P_1P_2)$ have been measured, and the data within $2\sigma$ errors are
\begin{eqnarray}
&&\mathcal{B}(D^+_s\to f_0(980)e^+\nu_e,~f_0(980)\to \pi^+\pi^-)= (1.30\pm0.63) \times10^{-3}  ~~\mbox{\cite{Hietala:2015jqa}},\nonumber\\
&&\mathcal{B}(D^+_s\to f_0(980)e^+\nu_e,~f_0(980)\to \pi^0\pi^0)= (7.9\pm2.9)\times10^{-4} ~~\mbox{\cite{BESIII:2021pdt}},\nonumber\\
&&\mathcal{B}(D^0\to a_0(980)^-e^+\nu_e,~a_0(980)^-\to \eta\pi^-)=  (1.33^{+0.68}_{-0.60})\times10^{-4}~~\mbox{ \cite{PDG2022}},\nonumber\\
&&\mathcal{B}(D^+\to a_0(980)^0e^+\nu_e,~a_0(980)^0\to \eta\pi^0)= (1.7^{+1.6}_{-1.4})\times10^{-4}~~\mbox{ \cite{PDG2022}},\nonumber\\
&&\mathcal{B}(D^+\to f_0(500)e^+\nu_e,~f_0(500)\to \pi^+\pi^-)= (6.3\pm1.0)\times10^{-4}~~\mbox{ \cite{PDG2022}}, \nonumber\\
&&\mathcal{B}(D^+\to f_0(980)e^+\nu_e,~f_0(980)\to \pi^+\pi^-) <2.8\times10^{-5}~~\mbox{ \cite{BESIII:2018qmf}},\nonumber\\
&&\mathcal{B}(D^+_s\to f_0(500)e^+\nu_e,~f_0(500)\to \pi^0\pi^0) <6.4\times10^{-4}~~ \mbox{\cite{BESIII:2021pdt}}.  \label{Eq:BrD2SlvSPP}
\end{eqnarray}

Two cases $S_{1}$ and $S_{2}$ will be considered  in the  $D\rightarrow S\ell^+\nu_\ell$  decays.  In $S_1$ case,  only experimental datum of $\mathcal{B}(D^+_s\to f_0(980)e^+\nu_e)$ is used to constrain one parameter  $c^S_0$  or  $c'^S_0$ and then predict other not yet measured  branching ratios.   The numerical results of $\mathcal{B}(D\to S\ell^+\nu)$ in $S_1$ case are given in the 2-4th and 8th columns of Tab. \ref{Tab:BrD2Slv}.
In the $S_2$ case,  the experimental data of both  $\mathcal{B}(D^+_s\to f_0(980)e^+\nu_e)$ in Eq. (\ref{Eq:BrDs2f0980ev}) and $\mathcal{B}(D\to S\ell^+\nu_\ell,~S\to P_1P_2)$ in Eq. (\ref{Eq:BrD2SlvSPP}) will be used to constrain the parameter  $c^S_0$  or  $c'^S_0$.
 The  predictions of $\mathcal{B}(D\to S\ell^+\nu)$ in $S_2$ case are listed in the 5-7th and 9th columns of Tab. \ref{Tab:BrD2Slv}.   Our comments on the results in the $S_{1,2}$ cases are as follows.
\begin{itemize}
\item {\bf Results in the two quark picture:}
In the two quark picture,  the three possible ranges of the mixing angle, $25^\circ<\theta_S<40^\circ$, $140^\circ<\theta_S<165^\circ $ and $-30^\circ<\theta_S<30^\circ $ \cite{Cheng:2005nb,LHCb:2013dkk} have been analyzed.
In $S_1$ case,  using the data of $\mathcal{B}(D^+_s\to f_0(980)e^+\nu_e)$, many predictions of $\mathcal{B}(D\to S\ell^+\nu)$ are obtained. As given in  the 2-4th columns of  Tab. \ref{Tab:BrD2Slv}, one can see that the predictions with $25^\circ<\theta_S<40^\circ$ are similar to ones with $140^\circ<\theta_S<165^\circ $,   the predictions  with $-30^\circ<\theta_S<30^\circ$ are slightly different from the first two, and the errors of predictions are quite large.
After adding the  experimental bounds of $\mathcal{B}(D\to S\ell^+\nu_\ell,~S\to P_1P_2)$, as given in the 5-7th columns of  Tab. \ref{Tab:BrD2Slv}, the three possible ranges of the mixing angle $\theta_S$ are obviously constrained, and they reduce to  $25^\circ<\theta_S<35^\circ$,    $144^\circ<\theta_S<158^\circ$  and  $22^\circ\leq|\theta_S|\leq30^\circ$, respectively.  In addition, the error of every prediction become smaller   by adding the  experimental bounds of $\mathcal{B}(D\to S\ell^+\nu_\ell,~S\to P_1P_2)$.

\item {\bf Results in the four quark picture:}
The predictions in the four quark picture are listed in the 8-9th columns of Tab. \ref{Tab:BrD2Slv}.
The majority of predictions in   four quark picture are smaller than  corresponding ones in  two quark picture.
Strong coupling constants $g'_4$ and $g_4$ are appeared  in $S\to P_1P_2$ decays with the four quark picture of light scalar mesons. At present, we only can determine  $\big|\frac{g'_4}{g_4}\big|$ from the $S\to P_1P_2$ decays.  The results of involved decays  with  both $\frac{g'_4}{g_4}>0$ and  $\frac{g'_4}{g_4}<0$ are given in the 9th column of  Tab. \ref{Tab:BrD2Slv}, and one can see that, except $\mathcal{B}(D^+_s\to f_0(500)e^+\nu_e)$  and $\mathcal{B}(D^+_s\to f_0(980)\mu^+\nu_\mu)$,  the other involved branching ratios are not obviously affected by the choice of $\frac{g'_4}{g_4}>0$ or $\frac{g'_4}{g_4}<0$.
The errors of the branching ratio predictions are obviously  reduced by the  experimental bounds of $\mathcal{B}(D\to S\ell^+\nu_\ell,~S\to P_1P_2)$.

\item {\bf Comparing with previous predictions:}
Previous predictions are listed in the last column of Tab. \ref{Tab:BrD2Slv}.  $\mathcal{B}(D^+_s\to f_0(500)e^+\nu_e)$, $\mathcal{B}(D^+_s\to f_0(500)\mu^+\nu_\mu)$ and $\mathcal{B}(D^+\to f_0(500)\mu^+\nu_\mu)$ are predicted for the first time.  Our predictions of
$\mathcal{B}(D^+_s\to f_0(980)\mu^+\nu_\mu)$, $\mathcal{B}(D^+\to a^0_0e^+\nu_e)$, $\mathcal{B}(D^+\to f_0(980) e^+\nu_e)$, $\mathcal{B}(D^+\to f_0(500) e^+\nu_e)$ and $\mathcal{B}(D^+\to a^0_0\mu^+\nu_\mu)$ are consistent with previous predictions in Refs. \cite{Soni:2020sgn,Colangelo:2010bg,Wang:2009azc}.  Our other predictions are about one order smaller or one order larger than previous ones in Refs. \cite{Momeni:2022gqb,Cheng:2017fkw}.

\end{itemize}

\begin{sidewaystable}
\renewcommand\arraystretch{1.3}
\tabcolsep 0.05in
\centering
\caption{ Branching ratios of  $D\to S\ell^+\nu$ decays within $2\sigma$ errors. As given in Ref. \cite{wang:2022ourwork},  $g'_4$ and $g_4$ are strong coupling constants obtained by the SU(3) flavor symmetry in $S\to P_1P_2$ decays, $^a$denotes the results  with $\frac{g'_4}{g_4}>0$, and $^b$denotes ones with $\frac{g'_4}{g_4}<0$, $\dag$denotes the results with two quark picture, and $\ddag$denotes the results with four quark picture.}\vspace{0.1cm}
{\footnotesize\begin{tabular}{l|ccc|ccc|c|c|c}  \hline
 Branching ratios   &   \multicolumn{3}{c|}{ones for $2q$ state in $S_1$} & \multicolumn{3}{c|}{ones for $2q$ state in $S_2$ }& ones for $4q$ &  ones for $4q$ &Previous ones   \\
  &$[25^\circ,40^\circ]$  & $[140^\circ,165^\circ]$ & $[-30^\circ,30^\circ]$  &$[25^\circ,35^\circ]$  & $[144^\circ,158^\circ]$ & $22^\circ\leq|\theta_S|\leq30^\circ$&  state in $S_1$&state in $S_2$ \\\hline
$\mathcal{B}(D^0\to K^-_0e^+\nu_e)(\times10^{-3})$                       &$3.38\pm2.12 $  &$3.18\pm2.05$  &$2.57\pm1.58$      &$3.02\pm1.11$   &$3.00\pm1.10$  &$2.98\pm1.05$    &$1.11\pm0.63 $
   &$1.25\pm0.45$& $0.103\pm0.115^\dag$ \cite{Momeni:2022gqb}\\
$\mathcal{B}(D^+\to \overline{K}_0^0e^+\nu_e)(\times10^{-3})$            &$8.66\pm5.55 $  &$7.99\pm5.02$  &$7.02\pm4.48$      &$7.74\pm2.88$   &$7.78\pm2.77$  &$7.68\pm2.78$    &$2.85\pm1.65 $     &$3.36\pm1.25$& $38.8\pm5.6^\dag$ \cite{Momeni:2022gqb}\\
$\mathcal{B}(D^+_s\to f_0(980)e^+\nu_e)(\times10^{-3})$                  &$ 2.30\pm0.80$  &$2.30\pm0.80$  &$2.30\pm0.80$      &$2.58\pm0.52$   &$2.57\pm0.53$   &$2.71\pm0.39$   &$ 2.30\pm0.80$  &$^{2.49\pm0.61^a}_{2.54\pm0.56^b}$&$2.1\pm0.2^\dag~\mbox{\cite{Soni:2020sgn}},~2^{+0.5\dag}_{-0.4}~\mbox{\cite{Colangelo:2010bg}}$ \\
$\mathcal{B}(D^+_s\to f_0(500)e^+\nu_e)(\times10^{-3})$                  &$6.73\pm6.11$   &$5.98\pm5.75$  &$3.25\pm3.25$      &$1.49\pm0.43$   &$1.45\pm0.46$  &$1.42\pm0.50$    &$0.37\pm0.37 $  &$^{0.31\pm0.31^a}_{0.17\pm0.17^b}$\\
\hline
$\mathcal{B}(D^0\to K^-_0\mu^+\nu_\mu)(\times10^{-3})$                   &$ 2.90\pm1.84$  &$2.73\pm1.77$  &$2.20\pm1.36$      &$2.59\pm0.97$   &$2.57\pm0.96$  &$2.56\pm0.92$    &$0.95\pm0.54$      &$1.09\pm0.39$& $0.103\pm0.115^\dag$ \cite{Momeni:2022gqb}\\
$\mathcal{B}(D^+\to \overline{K}_0^0\mu^+\nu_\mu)(\times10^{-3})$        &$7.46\pm4.81 $  &$6.87\pm4.33$  &$6.04\pm3.88$      &$6.65\pm2.52$   &$6.69\pm2.43$  &$6.59\pm2.43$    &$2.45\pm1.43 $     &$2.89\pm1.09$&$38.8\pm5.6^\dag$ \cite{Momeni:2022gqb}\\
$\mathcal{B}(D^+_s\to f_0(980)\mu^+\nu_\mu)(\times10^{-3})$              &$1.95\pm0.70$   &$1.95\pm0.70$  &$1.95\pm0.69$      &$2.20\pm0.45$   &$2.20\pm0.45$  &$2.32\pm0.33$    &$1.95\pm0.70 $  &$^{2.12\pm0.54^a}_{2.16\pm0.49^b}$&$2.1\pm0.2^\dag$ \cite{Soni:2020sgn}\\
$\mathcal{B}(D^+_s\to f_0(500)\mu^+\nu_\mu)(\times10^{-3})$              &$6.21\pm5.66$   &$5.53\pm5.32$  &$3.01\pm3.01$      &$1.33\pm0.39$   &$1.31\pm0.43$  &$1.28\pm0.46$    &$0.34\pm0.34$   &$^{0.29\pm0.29^a}_{0.16\pm0.16^b}$\\
\hline
$\mathcal{B}(D^0\to a^-_0e^+\nu_e)(\times10^{-5})$                       &$9.99\pm6.54$   &$9.56\pm6.50$  &$8.34\pm5.67$      &$9.22\pm3.98$   &$9.09\pm3.65$  &$9.17\pm3.58$    &$3.42\pm2.06 $  &$4.32\pm1.17$&$^{16.8\pm1.5^\dag~\mbox{\cite{Soni:2020sgn}},~40.8^{+13.7\dag}_{-12.2}~\mbox{\cite{Cheng:2017fkw}},}_{ 24.4\pm3.0^\dag~\mbox{\cite{Momeni:2022gqb}}}$\\
$\mathcal{B}(D^+\to a^0_0e^+\nu_e)(\times10^{-5})$                      &$ 13.09\pm8.62$  &$12.62\pm8.67$  &$10.89\pm7.35$    &$12.09\pm5.19$  &$11.81\pm4.71$  &$11.97\pm4.66$    &$4.49\pm2.71 $ &$5.68\pm1.52$&$^{21.8\pm3.8^\dag~\mbox{\cite{Soni:2020sgn}},~54.0^{+17.8\dag}_{-15.9}~\mbox{\cite{Cheng:2017fkw}}}_{6\sim8^{\dag}\mbox{\cite{Wang:2009azc}},~5\sim5.4^{\ddag}\mbox{\cite{Wang:2009azc}}}$ \\                                                                                                 %
$\mathcal{B}(D^+\to f_0(980) e^+\nu_e)(\times10^{-5})$                  &$3.92\pm2.92$    &$3.48\pm3.13$  &$1.59\pm1.59$      &$2.62\pm0.82$   &$2.52\pm0.94$  &$2.40\pm0.80$    &$3.14\pm1.98 $  &$^{3.35\pm1.80^a}_{3.89\pm1.35^b}$ &$^{7.78\pm0.68^\dag~\mbox{\cite{Soni:2020sgn}},~5.7\pm1.3^\dag~\mbox{\cite{Ke:2009ed}}}_{0.4\sim3.5^\dag\mbox{\cite{Wang:2009azc}},~1.9\sim6.3^\ddag\mbox{\cite{Wang:2009azc}}}$ \\                                                                                         %
$\mathcal{B}(D^+\to f_0(500) e^+\nu_e)(\times10^{-4})$                  &$4.05\pm3.20$    &$4.08\pm3.10$  &$4.21\pm3.28$      &$2.16\pm0.96$   &$2.59\pm1.38$  &$2.70\pm1.28$    &$ 4.97\pm4.13$  &$^{4.97\pm3.34^a}_{4.95\pm3.36^b}$&$0.4\sim0.6^\dag$\cite{Wang:2009azc},$~0.88\sim1.4^\ddag$\cite{Wang:2009azc} \\                                                                                                                                                             %
$\mathcal{B}(D^+_s\to K^0_0e^+\nu_e)(\times10^{-4})$                    &$3.73\pm2.37 $   &$3.41\pm2.13$  &$2.99\pm1.88$      &$3.35\pm1.21$    &$3.32\pm1.20$  &$3.35\pm1.15$   &$1.25\pm0.71$       &$1.43\pm0.51$&$26.5\pm2.8^\dag~\mbox{\cite{Momeni:2022gqb}}$  \\\hline
$\mathcal{B}(D^0\to a^-_0\mu^+\nu_\mu)(\times10^{-5})$                  &$8.25\pm5.45 $   &$7.89\pm5.42$  &$6.91\pm4.75$      &$7.61\pm3.37$   &$7.51\pm3.10$  &$7.57\pm3.04$    &$2.83\pm1.72 $      &$3.57\pm0.99$&$16.3\pm1.4^\dag$~\cite{Soni:2020sgn},~$24.4\pm3.0^\dag~\mbox{\cite{Momeni:2022gqb}}$\\
$\mathcal{B}(D^+\to a^0_0\mu^+\nu_\mu)(\times10^{-5})$                  &$10.83\pm7.19 $  &$10.44\pm7.23$  &$9.04\pm6.16$     &$10.00\pm4.41$   &$9.76\pm4.00$  &$9.89\pm3.97$    &$ 3.73\pm2.28$ &$4.69\pm1.30$&$21.2\pm3.7^\dag$~\cite{Soni:2020sgn}\\
$\mathcal{B}(D^+\to f_0(980) \mu^+\nu_\mu)(\times10^{-5})$              &$3.23\pm2.41$    &$2.88\pm2.60$  &$1.32\pm1.32$     &$2.15\pm0.70$   &$2.09\pm0.78$  &$1.99\pm0.66$     &$ 2.56\pm1.62$  &$^{2.74\pm1.49^a}_{3.20\pm1.14^b}$&$7.87\pm0.67^\dag$~\cite{Soni:2020sgn} \\
$\mathcal{B}(D^+\to f_0(500) \mu^+\nu_\mu)(\times10^{-4})$              &$3.69\pm2.96$    &$3.71\pm2.86$  &$3.84\pm3.04$     &$1.92\pm0.88$   &$2.32\pm1.27$  &$2.42\pm1.19$     &$4.54\pm3.81 $  &$^{4.52\pm3.10^a}_{4.49\pm3.12^b}$\\
$\mathcal{B}(D^+_s\to K^0_0\mu^+\nu_\mu)(\times10^{-4})$                &$3.28\pm2.10 $   &$3.00\pm1.88$  &$2.62\pm1.66$     &$2.94\pm1.08$   &$2.91\pm1.06$  &$2.94\pm1.02$     &$ 1.10\pm0.63$      &$1.26\pm0.45$&$26.5\pm2.8^\dag~\mbox{\cite{Momeni:2022gqb}}$ \\\hline
\end{tabular}}\label{Tab:BrD2Slv}
\end{sidewaystable}

\newpage

\section{Summary}
Many semileptonic $D \to P/V/S\ell^+\nu_\ell$ decays have been measured, and these processes could be used to test the SU(3) flavor symmetry approach.
In terms of  the SU(3) flavor symmetry and the SU(3) flavor breaking,   the amplitude relations   have been obtained.
Then using the present data of $\mathcal{B}(D \to P/V/S\ell^+\nu_\ell)$, we have presented a theoretical analysis of the $D \to P/V/S\ell^+\nu_\ell$ decays.
Our main results can be summarized as follows.
\begin{itemize}
\item {\bf $D\to P\ell^+\nu_\ell$ decays:} Our predictions with the SU(3) flavor symmetry in the $C_1$ case  and the predictions after adding  SU(3) flavor breaking contributions in the $C_4$ case are quite consistent with all present experimental data of $\mathcal{B}(D\to P\ell^+\nu_\ell)$  within $2\sigma$  errors.   In the $C_{2}$ and $C_{3}$  cases, our  SU(3) flavor symmetry predictions are consistent with all present experimental data except $\mathcal{B}(D^+\to \pi^0\ell^+\nu_\ell)$ and $\mathcal{B}(D^0\to \pi^-\ell^+\nu_\ell)$, which are slight larger than their experiential upper limits.  The not yet measured
 $\mathcal{B}(D^+_s\to \pi^0e^+\nu_e), ~\mathcal{B}(D^+\to \eta'\mu^+\nu_\mu),~ \mathcal{B}(D^+_s\to K^0\mu^+\nu_\mu),~\mathcal{B}(D^+_s\to \pi^0\mu^+\nu_\mu),~\mathcal{B}(D^+_s\to \pi^0\tau^+\nu_\tau)$ and the lepton flavor universality parameters have been obtained. Moreover, the forward-backward asymmetries, the lepton-side convexity parameters, the longitudinal (transverse) polarizations of   the final charged leptons   with two ways of integration for the  $D\to P\ell^+\nu_\ell$ decays have been predicted.  The    $q^2$ dependence of corresponding  differential quantities of the $D\to P\ell^+\nu_\ell$ decays  in the $C_3$ case  have been  displayed.

\item {\bf $D\to V\ell^+\nu_\ell$ decays:}  As given in the $C_1$, $C_2$ and $C_3$ cases, our  SU(3) flavor symmetry  predictions of $\mathcal{B}(D^+\to \omega e^+\nu_e)$ and $\mathcal{B}(D^0\to \rho^-\mu^+\nu_\mu)$ are slightly larger than its experimental upper limits, and other  SU(3) flavor symmetry predictions are consistent with present data. After considering the SU(3) flavor breaking effects, as given in the $C_4$ case, all predictions are consistent with present data.   The not yet measured  or not yet well measured branching ratios of $D^+\to \phi e^+\nu_e,~D^+_s\to \omega e^+\nu_e,~D^+\to \phi \mu^+\nu_\mu,~D^+_s\to \omega \mu^+\nu_\mu,$ and $D^+_s\to K^{*0}\mu^+\nu_\mu$ have been predicted.   The    $q^2$ dependence of corresponding  differential quantities of the $D\to V\ell^+\nu_\ell$ decays in the $C_3$ case have also been  displayed.

\item {\bf $D\to S\ell^+\nu_\ell$ decays: }  Among 18 $D\to S\ell^+\nu_\ell$ decay modes, only $\mathcal{B}(D^+_s\to f_0(980)e^+\nu_e)$  has been measured, and  this experimental datum has been  used to constrain the  SU(3) flavor symmetry parameter and then predict other not yet measured  branching ratios. Furthermore,  the relevant experimental bounds of $\mathcal{B}(D\to S\ell^+\nu_\ell,~S\to P_1P_2)$ have also been added.  The two quark and the four quark scenarios for the light scalar mesons  are considered, and   the three possible ranges of the mixing angle $\theta_S$ in the two quark picture have been analyzed.

\end{itemize}

The SU(3) flavor  symmetry is approximate approach, and it can
still provide  very useful information.  We have found that the SU(3) flavor symmetry approach works well in the semileptonic $D \to P/V\ell^+\nu_\ell$ decays, and the SU(3) flavor symmetry predictions of the $D \to S\ell^+\nu_\ell$ decays need to be  further tested, and our predictions of the $D \to S\ell^+\nu_\ell$ decays are useful for  probing the structure of light scalar mesons.
 According to our  predictions, some decay modes could be observed at BESIII,  LHCb or BelleII   in near future experiments.

\section*{ACKNOWLEDGEMENTS}
The work was supported by the National Natural Science Foundation of China (12175088).

\section*{References}


\begin{thebibliography}{99}

\bibitem{PDG2022}
R. L. Workman et al. (Particle Data Group), Prog. Theor. Exp. Phys. 2022, 083C01 (2022).


\bibitem{Melikhov:2000yu}
D.~Melikhov and B.~Stech,
Phys. Rev. D \textbf{62} (2000), 014006
[arXiv:hep-ph/0001113 [hep-ph]].

\bibitem{Cheng:2017pcq}
H.~Y.~Cheng and X.~W.~Kang,
Eur. Phys. J. C \textbf{77} (2017) no.9, 587
[erratum: Eur. Phys. J. C \textbf{77} (2017) no.12, 863]
[arXiv:1707.02851 [hep-ph]].


\bibitem{Soni:2018adu}
N.~R.~Soni, M.~A.~Ivanov, J.~G.~K\"orner, J.~N.~Pandya, P.~Santorelli and C.~T.~Tran,
Phys. Rev. D \textbf{98} (2018) no.11, 114031
[arXiv:1810.11907 [hep-ph]].



\bibitem{Chang:2018zjq}
Q.~Chang, X.~N.~Li, X.~Q.~Li, F.~Su and Y.~D.~Yang,
Phys. Rev. D \textbf{98} (2018) no.11, 114018
[arXiv:1810.00296 [hep-ph]].


\bibitem{Chang:2020wvs}
Q.~Chang, X.~L.~Wang and L.~T.~Wang,
Chin. Phys. C \textbf{44} (2020) no.8, 083105
[arXiv:2003.10833 [hep-ph]].


\bibitem{Faustov:2019mqr}
R.~N.~Faustov, V.~O.~Galkin and X.~W.~Kang,
Phys. Rev. D \textbf{101} (2020) no.1, 013004
[arXiv:1911.08209 [hep-ph]].


\bibitem{Ball:1993tp}
P.~Ball,
Phys. Rev. D \textbf{48} (1993), 3190-3203
[arXiv:hep-ph/9305267 [hep-ph]].


\bibitem{Bhattacharyya:2017wxk}
S.~Bhattacharyya, M.~Haiduc, A.~Tania Neagu and E.~Firu,
Eur. Phys. J. Plus \textbf{134} (2019) no.1, 37
[arXiv:1709.00882 [nucl-ex]].

\bibitem{Fu:2018yin}
H.~B.~Fu, L.~Zeng, R.~L\"u, W.~Cheng and X.~G.~Wu,
Eur. Phys. J. C \textbf{80} (2020) no.3, 194
[arXiv:1808.06412 [hep-ph]].


\bibitem{Fu:2020vqd}
H.~B.~Fu, W.~Cheng, L.~Zeng, D.~D.~Hu and T.~Zhong,
Phys. Rev. Res. \textbf{2} (2020) no.4, 043129
[arXiv:2003.07626 [hep-ph]].


\bibitem{Grach:1996nz}
I.~L.~Grach, I.~M.~Narodetsky and S.~Simula,
Phys. Lett. B \textbf{385} (1996), 317-323
[arXiv:hep-ph/9605349 [hep-ph]].

\bibitem{Cheng:2003sm}
H.~Y.~Cheng, C.~K.~Chua and C.~W.~Hwang,
Phys. Rev. D \textbf{69} (2004), 074025
[arXiv:hep-ph/0310359 [hep-ph]].

\bibitem{Chang:2019mmh}
Q.~Chang, X.~N.~Li and L.~T.~Wang,
Eur. Phys. J. C \textbf{79} (2019) no.5, 422
[arXiv:1905.05098 [hep-ph]].


\bibitem{Lubicz:2018rfs}
V.~Lubicz \textit{et al.} [ETM],
Phys. Rev. D \textbf{98} (2018) no.1, 014516
[arXiv:1803.04807 [hep-lat]].


\bibitem{Lubicz:2017syv}
V.~Lubicz \textit{et al.} [ETM],
Phys. Rev. D \textbf{96} (2017) no.5, 054514
[erratum: Phys. Rev. D \textbf{99} (2019) no.9, 099902; erratum: Phys. Rev. D \textbf{100} (2019) no.7, 079901]
[arXiv:1706.03017 [hep-lat]].

\bibitem{He:1998rq}
  X.~G.~He,
  Eur.\ Phys.\ J.\ C {\bf 9}, 443 (1999)
  [hep-ph/9810397].

\bibitem{He:2000ys}
  X.~G.~He, Y.~K.~Hsiao, J.~Q.~Shi, Y.~L.~Wu and Y.~F.~Zhou,
  Phys.\ Rev.\ D {\bf 64}, 034002 (2001)
  [hep-ph/0011337].


\bibitem{Fu:2003fy}
  H.~K.~Fu, X.~G.~He and Y.~K.~Hsiao,
  Phys.\ Rev.\ D {\bf 69}, 074002 (2004)
  [hep-ph/0304242].

\bibitem{Hsiao:2015iiu}
  Y.~K.~Hsiao, C.~F.~Chang and X.~G.~He,
  Phys.\ Rev.\ D {\bf 93}, no. 11, 114002 (2016)
  [arXiv:1512.09223 [hep-ph]].

\bibitem{He:2015fwa}
  X.~G.~He and G.~N.~Li,
  Phys.\ Lett.\ B {\bf 750}, 82 (2015)
  [arXiv:1501.00646 [hep-ph]].

\bibitem{Gronau:1994rj}
  M.~Gronau, O.~F.~Hernandez, D.~London and J.~L.~Rosner,
  Phys.\ Rev.\ D {\bf 50}, 4529 (1994)
  [hep-ph/9404283].

\bibitem{Gronau:1995hm}
  M.~Gronau, O.~F.~Hernandez, D.~London and J.~L.~Rosner,
  Phys.\ Rev.\ D {\bf 52}, 6356 (1995)
  [hep-ph/9504326].

\bibitem{Zhou:2016jkv}
  S.~H.~Zhou, Q.~A.~Zhang, W.~R.~Lyu and C.~D.~L\"{u},
  Eur.\ Phys.\ J.\ C {\bf 77}, no. 2, 125 (2017)
  [arXiv:1608.02819 [hep-ph]].


\bibitem{Cheng:2014rfa}
  H.~Y.~Cheng, C.~W.~Chiang and A.~L.~Kuo,
  Phys.\ Rev.\ D {\bf 91}, no. 1, 014011 (2015)
  [arXiv:1409.5026 [hep-ph]].



\bibitem{He:2015fsa}
  M.~He, X.~G.~He and G.~N.~Li,
  Phys.\ Rev.\ D {\bf 92}, no. 3, 036010 (2015)
  [arXiv:1507.07990 [hep-ph]].

 \bibitem{Deshpande:1994ii}
  N.~G.~Deshpande and X.~G.~He,
  Phys.\ Rev.\ Lett.\  {\bf 75}, 1703 (1995)
  [hep-ph/9412393].

  \bibitem{Shivashankara:2015cta}
  S.~Shivashankara, W.~Wu and A.~Datta,
  Phys.\ Rev.\ D {\bf 91},  115003 (2015)
  [arXiv:1502.07230 [hep-ph]].

\bibitem{Wang:2021uzi}
R.~M.~Wang, Y.~G.~Xu, C.~Hua and X.~D.~Cheng,
Phys. Rev. D \textbf{103} (2021) no.1, 013007
[arXiv:2101.02421 [hep-ph]].

\bibitem{Wang:2020wxn}
R.~M.~Wang, X.~D.~Cheng, Y.~Y.~Fan, J.~L.~Zhang and Y.~G.~Xu,
J. Phys. G \textbf{48} (2021) no.8, 085001
[arXiv:2008.06624 [hep-ph]].




  \bibitem{Grossman:2012ry}
  Y.~Grossman and D.~J.~Robinson,
  JHEP {\bf 1304}, 067 (2013)
  [arXiv:1211.3361 [hep-ph]].


  \bibitem{Pirtskhalava:2011va}
  D.~Pirtskhalava and P.~Uttayarat,
  Phys.\ Lett.\ B {\bf 712}, 81 (2012)
  [arXiv:1112.5451 [hep-ph]].


\bibitem{Cheng:2012xb}
  H.~Y.~Cheng and C.~W.~Chiang,
  Phys.\ Rev.\ D {\bf 86}, 014014 (2012)
  [arXiv:1205.0580 [hep-ph]].

\bibitem{Savage:1989qr}
  M.~J.~Savage and R.~P.~Springer,
  Phys.\ Rev.\ D {\bf 42}, 1527 (1990).


\bibitem{Savage:1991wu}
  M.~J.~Savage,
  Phys.\ Lett.\ B {\bf 257}, 414 (1991).



\bibitem{Altarelli:1975ye}
  G.~Altarelli, N.~Cabibbo and L.~Maiani,
  Phys.\ Lett.\  {\bf 57B}, 277 (1975).

\bibitem{Lu:2016ogy}
  C.~D.~L\"{u}, W.~Wang and F.~S.~Yu,
  Phys.\ Rev.\ D {\bf 93}, no. 5, 056008 (2016)
  [arXiv:1601.04241 [hep-ph]].


\bibitem{Geng:2017esc}
  C.~Q.~Geng, Y.~K.~Hsiao, Y.~H.~Lin and L.~L.~Liu,
  Phys.\ Lett.\ B {\bf 776}, 265 (2018)
  [arXiv:1708.02460 [hep-ph]].

\bibitem{Geng:2018plk}
  C.~Q.~Geng, Y.~K.~Hsiao, C.~W.~Liu and T.~H.~Tsai,
  Phys.\ Rev.\ D {\bf 97}, no. 7, 073006 (2018)
  [arXiv:1801.03276 [hep-ph]].

\bibitem{Geng:2017mxn}
  C.~Q.~Geng, Y.~K.~Hsiao, C.~W.~Liu and T.~H.~Tsai,
  JHEP {\bf 1711}, 147 (2017)
  [arXiv:1709.00808 [hep-ph]].

\bibitem{Geng:2019bfz}
  C.~Q.~Geng, C.~W.~Liu, T.~H.~Tsai and S.~W.~Yeh,
  arXiv:1901.05610 [hep-ph].


\bibitem{Wang:2017azm}
  W.~Wang, Z.~P.~Xing and J.~Xu,
  Eur.\ Phys.\ J.\ C {\bf 77}, no. 11, 800 (2017)
  [arXiv:1707.06570 [hep-ph]].


\bibitem{Wang:2019dls}
  D.~Wang,
  arXiv:1901.01776 [hep-ph].

\bibitem{Wang:2017gxe}
  D.~Wang, P.~F.~Guo, W.~H.~Long and F.~S.~Yu,
  JHEP {\bf 1803}, 066 (2018)
  [arXiv:1709.09873 [hep-ph]].


\bibitem{Muller:2015lua}
  S.~M\"{u}ller, U.~Nierste and S.~Schacht,
  Phys.\ Rev.\ D {\bf 92}, no. 1, 014004 (2015)
  [arXiv:1503.06759 [hep-ph]].




\bibitem{Barranco:2014bva}
  J.~Barranco, D.~Delepine, V.~Gonzalez Macias and L.~Lopez-Lozano,
  arXiv:1404.0454 [hep-ph].

\bibitem{Barranco:2013tba}
  J.~Barranco, D.~Delepine, V.~Gonzalez Macias and L.~Lopez-Lozano,
  Phys.\ Lett.\ B {\bf 731}, 36 (2014)
  [arXiv:1303.3896 [hep-ph]].

\bibitem{Akeroyd:2009tn}
  A.~G.~Akeroyd and F.~Mahmoudi,
  JHEP {\bf 0904}, 121 (2009)
  [arXiv:0902.2393 [hep-ph]].

\bibitem{Dobrescu:2008er}
  B.~A.~Dobrescu and A.~S.~Kronfeld,
  Phys.\ Rev.\ Lett.\  {\bf 100}, 241802 (2008)
  [arXiv:0803.0512 [hep-ph]].

\bibitem{Akeroyd:2007eh}
  A.~G.~Akeroyd and C.~H.~Chen,
  Phys.\ Rev.\ D {\bf 75}, 075004 (2007)
  [hep-ph/0701078].

\bibitem{Fajfer:2006uy}
  S.~Fajfer and J.~F.~Kamenik,
  Phys.\ Rev.\ D {\bf 73}, 057503 (2006)
  [hep-ph/0601028].

\bibitem{Fajfer:2005ug}
  S.~Fajfer and J.~F.~Kamenik,
  Phys.\ Rev.\ D {\bf 72}, 034029 (2005)
  [hep-ph/0506051].


\bibitem{Fajfer:2004mv}
  S.~Fajfer and J.~F.~Kamenik,
  Phys.\ Rev.\ D {\bf 71}, 014020 (2005)
  [hep-ph/0412140].



\bibitem{Akeroyd:2003jb}
  A.~G.~Akeroyd,
  Prog.\ Theor.\ Phys.\  {\bf 111}, 295 (2004)
  [hep-ph/0308260].

\bibitem{Akeroyd:2002pi}
  A.~G.~Akeroyd and S.~Recksiegel,
  Phys.\ Lett.\ B {\bf 554}, 38 (2003)
  [hep-ph/0210376].


\bibitem{Ivanov:2019nqd}
M.~A.~Ivanov, J.~G.~K\"orner, J.~N.~Pandya, P.~Santorelli, N.~R.~Soni and C.~T.~Tran,
Front. Phys. (Beijing) \textbf{14} (2019) no.6, 64401
[arXiv:1904.07740 [hep-ph]].

\bibitem{He:2018joe}
X.~G.~He, Y.~J.~Shi and W.~Wang,
Eur. Phys. J. C \textbf{80}, no.5, 359 (2020)
[arXiv:1811.03480 [hep-ph]].


\bibitem{Cheng:2005nb}
H.~Y.~Cheng, C.~K.~Chua and K.~C.~Yang,
Phys. Rev. D \textbf{73} (2006), 014017
[arXiv:hep-ph/0508104 [hep-ph]].


\bibitem{Maiani:2004uc}
L.~Maiani, F.~Piccinini, A.~D.~Polosa and V.~Riquer,
Phys. Rev. Lett. \textbf{93} (2004), 212002
[arXiv:hep-ph/0407017 [hep-ph]].



\bibitem{Dai:2018fmx}
L.~Y.~Dai, X.~W.~Kang and U.~G.~Mei\ss{}ner,
Phys. Rev. D \textbf{98} (2018) no.7, 074033
[arXiv:1808.05057 [hep-ph]].

\bibitem{tHooft:2008rus}
G.~'t Hooft, G.~Isidori, L.~Maiani, A.~D.~Polosa and V.~Riquer,
Phys. Lett. B \textbf{662} (2008), 424-430
[arXiv:0801.2288 [hep-ph]].

\bibitem{Pelaez:2003dy}
J.~R.~Pelaez,
Phys. Rev. Lett. \textbf{92} (2004), 102001
[arXiv:hep-ph/0309292 [hep-ph]].

\bibitem{Sun:2010nv}
Y.~J.~Sun, Z.~H.~Li and T.~Huang,
Phys. Rev. D \textbf{83} (2011), 025024
[arXiv:1011.3901 [hep-ph]].

\bibitem{Oller:1997ti}
J.~A.~Oller and E.~Oset,
Nucl. Phys. A \textbf{620} (1997), 438-456
[erratum: Nucl. Phys. A \textbf{652} (1999), 407-409]
[arXiv:hep-ph/9702314 [hep-ph]].


\bibitem{Baru:2003qq}
V.~Baru, J.~Haidenbauer, C.~Hanhart, Y.~Kalashnikova and A.~E.~Kudryavtsev,
Phys. Lett. B \textbf{586} (2004), 53-61
[arXiv:hep-ph/0308129 [hep-ph]].


\bibitem{Achasov:1996ei}
N.~N.~Achasov, V.~V.~Gubin and V.~I.~Shevchenko,
Phys. Rev. D \textbf{56} (1997), 203-211
[arXiv:hep-ph/9605245 [hep-ph]].



\bibitem{Momeni:2022gqb}
S.~Momeni and M.~Saghebfar,
Eur. Phys. J. C \textbf{82} (2022) no.5, 473


\bibitem{LHCb:2013dkk}
R.~Aaij \textit{et al.} [LHCb],
Phys. Rev. D \textbf{87} (2013) no.5, 052001
[arXiv:1301.5347 [hep-ex]].




\bibitem{Jaffe:1976ig}
R.~L.~Jaffe,
Phys. Rev. D \textbf{15} (1977), 267


\bibitem{Xu:2013dta}
D.~Xu, G.~N.~Li and X.~G.~He,
Int. J. Mod. Phys. A \textbf{29} (2014), 1450011
[arXiv:1307.7186 [hep-ph]].


\bibitem{Tanaka:2010se}
  M.~Tanaka and R.~Watanabe,
  Phys.\ Rev.\ D {\bf 82}, 034027 (2010)
  [arXiv:1005.4306 [hep-ph]].

\bibitem{Tanaka:1994ay}
  M.~Tanaka,
  Z.\ Phys.\ C {\bf 67}, 321 (1995)
  [hep-ph/9411405].

\bibitem{Fajfer:2012vx}
  S.~Fajfer, J.~F.~Kamenik and I.~Nisandzic,
  Phys.\ Rev.\ D {\bf 85}, 094025 (2012)
  [arXiv:1203.2654 [hep-ph]].



  \bibitem{Celis:2012dk}
  A.~Celis, M.~Jung, X.~Q.~Li and A.~Pich,
  JHEP {\bf 1301}, 054 (2013)
  [arXiv:1210.8443 [hep-ph]].


\bibitem{Bobeth:2010wg}
  C.~Bobeth, G.~Hiller and D.~van Dyk,
  JHEP {\bf 1007}, 098 (2010)
  [arXiv:1006.5013 [hep-ph]].


\bibitem{Li:2020ylu}
H.~B.~Li and M.~Z.~Yang,
Phys. Lett. B \textbf{811} (2020), 135879
[arXiv:2006.15798 [hep-ph]].


\bibitem{Achasov:2012kk}
N.~N.~Achasov and A.~V.~Kiselev,
Phys. Rev. D \textbf{86} (2012), 114010
[arXiv:1206.5500 [hep-ph]].


\bibitem{Soni:2020sgn}
N.~R.~Soni, A.~N.~Gadaria, J.~J.~Patel and J.~N.~Pandya,
Phys. Rev. D \textbf{102} (2020) no.1, 016013
[arXiv:2001.10195 [hep-ph]].


\bibitem{Leng:2020fei}
X.~Leng, X.~L.~Mu, Z.~T.~Zou and Y.~Li,
Chin. Phys. C \textbf{45} (2021) no.6, 063107
[arXiv:2011.01061 [hep-ph]].


\bibitem{wang:2022ourwork}
Ru-Min Wang, Yi-Jie Zhang, Yi Qiao, Xiao-Dong Cheng and Yuan-Guo Xu,
Four-body  Semileptonic Charm Decays  $D\to P_1P_2\ell^+\nu_\ell $  with the SU(3) Flavor Symmetry, in preparation.



\bibitem{Hietala:2015jqa}
J.~Hietala, D.~Cronin-Hennessy, T.~Pedlar and I.~Shipsey,
Phys. Rev. D \textbf{92} (2015) no.1, 012009
[arXiv:1505.04205 [hep-ex]].


\bibitem{BESIII:2021pdt}
M.~Ablikim \textit{et al.} [BESIII],
[arXiv:2110.13994 [hep-ex]].



\bibitem{BESIII:2018qmf}
M.~Ablikim \textit{et al.} [BESIII],
Phys. Rev. Lett. \textbf{122} (2019) no.6, 062001
[arXiv:1809.06496 [hep-ex]].




\bibitem{Colangelo:2010bg}
P.~Colangelo, F.~De Fazio and W.~Wang,
Phys. Rev. D \textbf{81} (2010), 074001
[arXiv:1002.2880 [hep-ph]].


\bibitem{Wang:2009azc}
W.~Wang and C.~D.~L\"{u},
Phys. Rev. D \textbf{82} (2010), 034016
[arXiv:0910.0613 [hep-ph]].



\bibitem{Cheng:2017fkw}
X.~D.~Cheng, H.~B.~Li, B.~Wei, Y.~G.~Xu and M.~Z.~Yang,
Phys. Rev. D \textbf{96} (2017) no.3, 033002
[arXiv:1706.01019 [hep-ph]].


\bibitem{Ke:2009ed}
H.~W.~Ke, X.~Q.~Li and Z.~T.~Wei,
Phys. Rev. D \textbf{80} (2009), 074030
[arXiv:0907.5465 [hep-ph]].



\end{thebibliography}
\end{document}